\documentclass[twocolumn,twocolappendix,trackchanges]{aastex701}
\usepackage{amsmath}
\usepackage{natbib}
\newcommand{\lbrac}[1]{\left(#1\right)}
\newcommand{\tc}{t_{\rm cool}}
\newcommand{\tcool}{\tc}
\newcommand{\tcz}{t_{\rm cool,ax}}
\newcommand{\tfz}{t_{\rm flow,ax}}
\newcommand{\tf}{t_{\rm flow}}

\newcommand{\nh}{n_{\rm H}}

\newcommand{\dd}{{\rm d}}
\newcommand{\regd}[2]{\frac{\dd #1}{\dd #2}}
\newcommand{\kms}{\,{\rm km\,s^{-1}}}
\newcommand{\Msunyr}{\,{\rm M_\odot\,yr^{-1}}}
\newcommand{\kpc}{\,{\rm kpc}}
\newcommand{\K}{\,{\rm K}}

\newcommand{\vrsph}{v_{r,\rm sph}}
\newcommand{\rhosph}{\rho_{\rm sph}}
\newcommand{\rhorcf}{\rho_{\rm rcf}}
\newcommand{\Tsph}{T_{\rm sph}}
\newcommand{\tcsph}{t_{\rm cool,sph}}
\newcommand{\vin}{v_{\rm in}}
\newcommand{\rin}{r_{\rm in}}
\newcommand{\rhoin}{\rho_{\rm in}}

\newcommand{\rhoax}{\rho_{\rm ax}}
\newcommand{\vrax}{v_{r,{\rm ax}}}
\newcommand{\Tax}{T_{\rm ax}}
\newcommand{\Kax}{K_{\rm ax}}
\newcommand{\nhax}{n_{\rm H,ax}}
\newcommand{\Pax}{P_{\rm ax}}
\newcommand{\csax}{c_{\rm s,ax}}
\newcommand{\emunits}{{\rm cm^{-6}\,pc}}
\newcommand{\vc}{v_{\rm c}}
\newcommand{\rout}{r_{\rm out}}
\newcommand{\risen}{r_{\rm isen}}

\newcommand{\vent}{v_{\rm ent}}

\begin{document}

\title{The hot circumgalactic medium: implications of mixing and resolution at the disk--halo interface}

\author[orcid=0000-0002-0311-2206]{Omry Ginzburg}
\affiliation{School of Physics \& Astronomy, Tel Aviv University, Tel Aviv 69978, Israel}
\email[show]{omrygin@tauex.tau.ac.il}  

\author[orcid=0000-0002-7541-9565]{Jonathan Stern} 
\affiliation{School of Physics \& Astronomy, Tel Aviv University, Tel Aviv 69978, Israel}
\email{sternnaty@gmail.com}

\author[orcid=0000-0003-3806-8548]{Drummond B. Fielding}
\affiliation{Department of Physics, New York University, 726 Broadway, New York, NY 10003, USA}
\email{dbf233@nyu.edu}

\author[orcid=0000-0002-4900-6628]{Claude-André Faucher-Giguère}
\affiliation{Center for Interdisciplinary Exploration and Research in Astrophysics (CIERA) and Department of Physics and Astronomy, Northwestern University, 1800
Sherman Ave, Evanston, IL 60201, USA}
\email{cgiguere@northwestern.edu}

\begin{abstract}
We demonstrate that microphysical processes at the disk--halo interface, typically unresolved in circumgalactic medium (CGM) simulations, strongly affect the thermodynamic structure and observational signatures of an inflowing hot ($\sim10^6\,{\rm K}$) CGM. If the hot gas is pulled through the interface faster than it can cool, e.g.\ by turbulent mixing with the ISM or fountain flows, the hot CGM approaches an isentropic state. If instead the interface is regulated by radiative cooling on the hot gas cooling time, the hot CGM develops an outward-increasing entropy profile akin to a classic cooling flow. The two regimes differ by a factor of $\sim\!10$ in the predicted hot gas density above the disk near the rotation axis, and consequently by a factor of $\sim100$ in the predicted X-ray emission measure. 
We show that an isentropic CGM is produced in idealized CGM simulations that fail to resolve the pc-scale disk--halo interface, as numerical mixing drives artificially rapid cooling at the  interface. A similar isentropic structure is found along the rotation axis in FIRE zoom-in simulations of $z\sim0$ Milky-Way-mass galaxies, where the cooling layer is likewise unresolved, raising the possibility that numerical mixing contributes to the isentropic structure.
In contrast, an outward-increasing entropy profile forms in our idealized simulations when we replace the unresolved interface with an inner boundary that enforces a slow inflow. This outward-increasing entropy profile is likely also present in the Milky Way hot CGM, inferred from comparing density estimates at $\sim100\,{\rm kpc}$ with the entropy of the eROSITA bubbles.

\end{abstract}



\defcitealias{Gupta23}{G23}
\section{Introduction}

Interfaces between gas phases with temperatures differing by orders of magnitude are ubiquitous in astrophysical systems. A prominent example occurs in the circumgalactic medium (CGM), where virial-temperature gas at $T\sim10^6\K$ can interface with cool gas at the photoionization equilibrium temperature of $T\sim10^4\K$. Such interfaces often mediate a mass flux from the hot gas into the cool gas (or vice versa), with a magnitude depending on physical processes operating at the interface. In the absence of strong mixing, hot gas cools radiatively on its own cooling timescale, producing a `cooling flow' \citep{Fabian84,Bertschinger89,McQuinnWerk18,Stern19,Hafen22,Sultan25}. Alternatively, turbulent mixing between the hot and cool phases can substantially shorten the effective cooling time and thereby increase the mass flux. Such turbulent radiative mixing layers (TRMLs) have been studied extensively in recent years in the CGM context \citep{Gronke18,Fielding20,Mandelker20}.

One particularly important hot--cool interface in the CGM is the disk--halo interface, where the cool ISM disk meets the hot CGM. Hot CGM atmospheres are expected to be prevalent around massive disk galaxies at low redshift, such as the Milky Way, since sufficiently long cooling times allow a quasi-hydrostatic hot CGM to form \citep{White78, White91, Birnboim03,Fielding17,Stern20}. The existence of a hot CGM is also supported by X-ray and UV absorption observations \citep[cf.\ \citealt{Salem16,Ji20}]{Spitzer56,Forman85,Li17,Faerman17,Faerman20,Bregman18,Stern19,Voit17}. Several mechanisms could drive turbulent mixing across the disk--halo interface and thus enhance the mass flux. These include galactic fountain flows, in which cool gas launched from the disk interacts and mixes with the inner hot CGM \citep[e.g.][]{Fraternali17,Li23}. Another possibility is a difference in rotation velocity between the ISM and hot CGM, with turbulence driven by the resulting shear \citep{sharma25,Lin25}. Estimates of the Milky-Way hot CGM rotation velocity of $\approx180\kms$ \citep{HodgesKluck16} are indeed lower than the ISM rotation velocity of $\approx240\kms$, however it is unclear whether this difference persists down to the interface since the hot CGM may spin up as it flows inwards \citep{Hafen22,Stern24,Sankar26}.

Previous studies of the disk--halo interface have largely focused either on its structure and observational signatures, or on the mass flux of fresh gas it can supply to the ISM \citep{FraternaliBinney08,Marasco12,Fraternali17}. Such accretion is required both to replenish gas consumed by star formation and to explain the chemical evolution of galaxies \citep[e.g.,][]{Melioli08}. In this context, turbulent mixing with fountain gas or due to shear has been proposed as a mechanism for enhancing the condensation rate of the hot CGM and thereby increasing the supply of gas to the disk \citep{Fraternali17, Lin25}. However, such an enhancement is not necessarily required, since cooling of the hot CGM on its own radiative cooling timescale can plausibly supply gas at a rate sufficient to sustain observed star formation rates \citep{Stern19,Stern20,Stern24}.

In this study we focus on implications of the disk-halo mass flux for the \textit{structure of the hot CGM} from which this gas is drawn. We demonstrate that the key distinction is whether the inflow timescale from the hot CGM to the disk-halo interface is shorter or comparable to the hot CGM radiative cooling time. If turbulent mixing at the disk--halo interface drives a sufficiently large mass flux, hot gas farther out flows inward faster than it can cool. Its entropy is therefore approximately conserved as it flows inward, yielding a nearly isentropic hot CGM and a relatively low density in the inner CGM. In contrast, if the mass flux through the interface is set by cooling of the hot CGM on its own cooling timescale, the gas loses substantial entropy as it flows inward, producing an entropy profile that increases outward as in a global cooling-flow solution\footnote{We distinguish between `local' cooling flows at hot-cool interfaces where the gas temperature actually drops \citep[e.g.][]{McQuinnWerk18}, and `global' cooling flows in which radiative losses are offset by compressive heating \citep[e.g.][]{Fabian84,Stern19}.} Thus, microphysical processes at the disk--halo interface, operating on scales typically far below the resolution of galaxy-scale and cosmological simulations, can determine the thermodynamic structure of the hot CGM on scales of tens of kiloparsecs.

This paper is organized as follows. In Section \ref{sec:analytics} we derive 1D solutions for a hot CGM inflow near the rotation axis, for different inner boundary conditions corresponding to different conditions at the disk-halo interface. In Section \ref{sec:idealized_simulations}, we test which solution emerges numerically in idealized simulations of hot and rotating CGM. In Section \ref{sec:fire_sims}, we analyze the flow near the axis in the CGM of Milky Way-mass galaxies in the FIRE-2 simulations. We then test the two solutions against observed estimates of the Milky Way CGM entropy profile in Section \ref{sec:mw_cgm}. We summarize our results in Section \ref{sec:summary}.

\section{Dependence of hot CGM solutions on inner boundary conditions}\label{sec:analytics}

To understand how the hot CGM structure depends on conditions at the disk-halo interface, we need to solve the flow equations from the $\sim\kpc$ scales of the interface to the $\approx10-100\kpc$ scales of the halo at which most hot CGM observations constrain (see below). A major difference between these scales is the importance of rotation -- at large radii centrifugal forces are small and the hot CGM is roughly spherical, 
while at inner radii rotation is significant and the hot CGM has more ellipsoidal constant-pressure curves \citep{Pezzulli17, Sormani19, Sobacchi19, Stern20, Stern24, Sankar26}. The inherent 2D nature of this (axisymmetric) problem makes finding full analytic solutions challenging, so for analytic tractability we focus on conditions near the rotation axis. Using simulations we show below that our conclusions apply also to off-axis regions, with the effect becoming weaker at larger angles from the rotation axis (see Section~\ref{sec:flowline_analysis}).

\subsection{Governing equations} 
\label{sec:governing_equations}

The hydrodynamic equations of a radiating ideal gas in an external gravitational potential, $\Phi$, are
\begin{equation}\label{eq:mass_full}
    \frac{\partial\rho}{\partial t} +\nabla\cdot\lbrac{\rho\vec{v}} = 0,
\end{equation}
\begin{equation}\label{eq:momentum_full}
    \frac{\partial \vec{v}}{\partial t}+\vec{v}\cdot\nabla\vec{v} = - \frac{1}{\rho}\nabla P - \nabla\Phi,
\end{equation}
\begin{equation}\label{eq:entropy_full}
    \frac{\partial \ln K}{\partial t} + \vec{v}\cdot\nabla\ln K = -\frac{1}{\tc},
\end{equation}
where $P,\rho$ and $\vec{v}$ are the pressure, density and velocity of the gas, respectively, $K = P\rho^{-5/3}$ is the entropy function (an adiabatic index $\gamma=5/3$ is assumed throughout), and $\tc$ is the cooling time, equal to
\begin{equation}
    \tc = \frac{3}{2}\frac{P}{\nh^2\Lambda},
\end{equation}
where $\nh$ is the hydrogen number density and $\Lambda$ is the cooling function, such that $\nh^2\Lambda$ is the energy lost per unit time per unit volume to radiation. Throughout the paper we use the cooling tables of \citet{Wiersma09}, assuming photoionization by the $z=0$ UV background from \citet{HaardtMadau}. 

Equation~\eqref{eq:entropy_full} neglects heating by supernova and AGN feedback, which can potentially balance radiative losses and sustain static or outflowing hot CGM \citep{Sharma12b,Voit17,Pezzulli17,Faerman17,Faerman20,Oppenheimer18, Carr22}. There is strong evidence that such heating exists in the ICM (the `cooling flow problem', e.g., \citealt{McDonald18}), though in the CGM inflow rates suggested by the X-ray luminosity are consistent with the SFR, and hence feedback heating may be small \citep{Stern19}. Weak heating by stellar processes is also suggested by the low velocity of observed outflows in these systems
\citep{Chen10,ThompsonHeckman, Popesso23}, though the Fermi bubbles could indicate more substantial heating \citep{Sarkar24}. In general, 
the level of AGN heating still has significant uncertainties and could be intermittent, in which case our neglect of the heating term would be valid whenever recurrence intervals exceed cooling times \citep{Stern19}. 

We can simplify eqns.~\eqref{eq:mass_full} -- \eqref{eq:entropy_full} by assuming that the hot CGM is `quasi-hydrostatic' with pressure-gradients and centrifugal forces roughly balancing gravity, 
i.e., with subsonic radial and polar velocities. 
Further assuming a steady-state flow and that the rotation axis is the $z$-axis, we get
\begin{equation}\label{eq:mass_axisymmetry}
    \frac{1}{r^2}\frac{\partial}{\partial  r}\lbrac{\rho v_rr^2} + \frac{1}{r\sin\theta}\frac{\partial}{\partial \theta}\lbrac{\rho v_\theta\sin\theta} = 0,
\end{equation}
\begin{equation}\label{eq:momentum_r_axisymmetry}
    \frac{\partial P}{\partial r} = -\rho\frac{\partial\Phi}{\partial r} + \rho\frac{v_\phi^2}{r},
\end{equation}
\begin{equation}\label{eq:momentum_theta_axisymmetry}
    \frac{\partial P}{\partial\theta} = \rho\frac{v_\phi^2\cos\theta}{\sin\theta} - \rho\frac{\partial \Phi}{\partial\theta},
\end{equation}
\begin{equation}\label{eq:momentum_phi_axisymmetry}
    v_r\frac{\partial}{\partial r}\lbrac{v_\phi r\sin\theta} + \frac{v_\theta}{r}\frac{\partial}{\partial\theta}\lbrac{v_\phi r\sin\theta} = 0,
\end{equation}
\begin{equation}\label{eq:entropy_axisymmetry}
    v_r\frac{\partial\ln K}{\partial r} + \frac{v_\theta}{r}\frac{\partial\ln K}{\partial\theta} = -\frac{1}{\tc}.
\end{equation}
where $v_r, v_\theta, v_\phi$ are the spherical components of the velocity, and we have neglected terms of order $v_r^2,v_\theta^2,v_rv_\theta$ due to the quasi-hydrostatic assumption. 

A mass sink at the disk-halo interface will cause the CGM to flow inward, 
so we search for an inflow solution. As the gas flows inwards,  conservation of angular momentum $j=rv_\phi\sin\theta$ along streamlines (eqn.~\ref{eq:momentum_phi_axisymmetry}) implies that the rotation velocity increases. This defines a characteristic `circularization radius' $R_c$ where $v_\phi\approx\vc$, with $v_c^2 = r\partial\Phi/\partial r$. 
For a Milky Way-mass hot CGM with spin parameter $\approx0.05$ we expect $R_c\approx15\kpc$. For $r^2\gg R_c^2$, centrifugal forces are small and thus we expect the flow to be approximately spherical with radial flowlines, and therefore $j$ is independent of $r$. Further assuming the rotation frequency at these radii is independent of $\theta$, as implied by non-radiative cosmological simulations \citep{Sharma12}, angular momentum conservation  implies
\begin{equation}\label{eq:am_profile}
    j(r\gg R_c,\theta)\approx R_cv_c\sin^2\theta.
\end{equation}

This limit of $r^2\gg R_c^2$, named the `slow rotation limit', allows an analytic solution to eqns.~\eqref{eq:mass_axisymmetry}--\eqref{eq:entropy_axisymmetry}, as found by \cite{Stern24}. Here we instead focus on smaller radii near the disk-halo interface where centrifugal forces are substantial. 

\subsection{Near-axis solutions}
To obtain solutions to eqns.~(\ref{eq:mass_axisymmetry})--(\ref{eq:entropy_axisymmetry}) which extend to $r^2\ll R_c^2$, we focus on the flow near the axis and expand the hydro and thermodynamic variables in powers of $\theta$. From symmetry and continuity $v_\theta$ and $v_\phi$ have only odd powers of $\theta$ in their expansion\footnote{By symmetry, $v_\theta$ and $v_\phi$ must go to zero as $\theta\to0$. The oddness of the powers is required in order to keep the derivatives finite at the axis.} while $v_r,P,\rho$ have only even powers. We therefore write
\begin{equation}\label{eq:expansion_theta}
    \begin{split}
        & P = \Pax(r)\lbrac{1+F_P(r)\theta^2+\ldots}, \\
        & \rho = \rhoax(r)\lbrac{1+F_\rho(r)\theta^2+\ldots}, \\
        & v_r = \vrax(r)\lbrac{1+F_r(r)\theta^2+\ldots}, \\
        & v_\theta = -\vrax F_\theta(r)\theta+\ldots, \\
        & v_\phi = v_c(R_c)F_\phi(r)\theta+\ldots,
    \end{split}
\end{equation}
where a subscript `ax' denotes values at the rotation axis. 
Plugging this approximation into eqns.~\eqref{eq:mass_axisymmetry} -- \eqref{eq:entropy_axisymmetry}, and dropping terms of order $\theta^2$ and above, we get the following equations for $v_r,P,\rho$ along the axis:
\begin{equation}\label{eq:mass_naa}
    \frac{\dd\ln\rhoax\vrax}{\dd\ln{r}} = -2\lbrac{1-F_\theta},
\end{equation}
\begin{equation}\label{eq:momentum_naa}
    \frac{\dd \Pax}{\dd r} = -\rhoax\frac{v_c^2}{r},
\end{equation}
\begin{equation}\label{eq:entropy_naa}
    \frac{\dd\ln{\Kax}}{\dd\ln {r}} = \frac{\tfz}{\tcz},
\end{equation}
where $\tfz\!=\!-r/\vrax$ and $\tcz$ are the flow time and cooling time along the axis, respectively. Equations~\eqref{eq:momentum_naa} and \eqref{eq:entropy_naa} are identical to the non-rotating case (Appendix~\ref{sec:spherical_solutions}), while the effect of rotation enters through mass conservation via $F_\theta$ (eqn.~\ref{eq:mass_naa}), which encodes the deflection of mass from the axis by the centrifugal force. 
For $F_\theta = 0$ the flow is spherical and the mass flux is $\rhoax\vrax\propto r^{-2}$, while for $F_\theta = 1$ the flow is vertical and $\rhoax \vrax$ is constant.  We therefore expect $F_\theta$ to rise from approximately zero at $r\gg R_c$, to order unity at $r\ll R_c$.  

An exact solution for $F_\theta$ needed for closing eqns.~(\ref{eq:mass_naa})--(\ref{eq:entropy_naa}) requires solving the higher order terms in $\theta$, which in turn requires a global solution since the flow is subsonic. We therefore instead parameterize $F_\theta$ using the $r\gg R_c$ limit of $F_\theta\sim 5R_c^2/9r^2$ \citep[their eqn.~41]{Stern24} 
\begin{equation}\label{eq:fiducia_ftheta}
    F_\theta(r) = \frac{F_0}{1+9F_0/5(r/R_c)^2},
\end{equation}
where $F_0$ is an adjustable parameter equal to $F_\theta$ at $r\!\ll\!R_c$. The idealized simulations below suggest $F_0\!\sim\!1$.

\subsection{Analytic approximations}

Below we integrate eqns.~\eqref{eq:mass_naa}--\eqref{eq:entropy_naa} numerically from $\rin\ll R_c$ to $\rout\gg R_c$. Intuition can however be gained also from analytic solutions, derivable at $r^2\ll R_c^2$ assuming a vertical flow $F_\theta=F_0=1$ and an isothermal potential (constant $\vc$). We use an inner boundary condition for velocity
\begin{equation}
    \vent\equiv -\vrax(\rin)~,
\end{equation}
which reflects the conditions at the disk-halo interface. We also  use an outer boundary condition for the density at the circularization radius $\rhoax(R_c)$ based on extrapolating the slow rotating solution down to this radius. 
Analytic solutions are found in two limits as described in Appendix~\ref{sec:analytic_solutions}. In one limit $\vent$ is set by the hot gas cooling time at the interface, and is equal to $3\,\rin/\tcool(\rin)$. Such solutions are termed \textit{cooling flow solutions}, and they result in
\begin{equation}\label{eq:tf_over_tc}
\begin{split}
    \Tax & = 6\times10^6 v_{\rm c,200}^2\,{\rm K},\\
    \nhax &= \nh(R_c)\left(\frac{r}{R_c}\right)^{-1/2},\\
    \vrax &= -5\left(\frac{\nh(R_c)}{10^{-3}\,{\rm cm}^{-3}}\right)\left(\frac{r}{\rin}\right)^{1/2}\kms,
\end{split}
\end{equation}
where $\nh=0.75\rho/m_{\rm p}$, $v_{\rm c,200}=\vc/200\kms$, and the numerical values assume $R_c=15\kpc$ and $\rin=1\kpc$. 
In this limit the temperature is constant and the density increases inward as $r^{-1/2}$. 

In the other limit, processes at the disk-halo interface such as turbulent or numerical mixing increase $\vent$ above the cooling flow value, so $\vent\!\gg \rin/\tcz$ which yields a flat entropy profile (see eqn.~\ref{eq:entropy_naa}). These solutions are thus termed \textit{isentropic solutions}. 
We show below idealized simulations with numerical mixing at the disk-halo interface that results in $v_{\rm ent}\!\sim\!40\kms$, so they fall in this category. 
The solutions are
\begin{equation}\label{eq:isentropic_sols}
\begin{split}
    \Tax & = \Tax(R_c)f(r), \\
    \rhoax & = \rhoax(R_c)f^{3/2}(r),\\ 
    \vrax & = -\vent f^{-3/2}(r)f^{3/2}(\rin),
\end{split}
\end{equation}
with $f$ a function which increases logarithmically with decreasing radius
\begin{equation}
    f(r) = 1+\frac{2}{3}\frac{v_c^2}{\csax^2(R_c)}\ln\lbrac{\frac{R_c}{r}}~,
\end{equation}
where $\csax = 150\,(\Tax/10^6\,{\rm K})^{0.5}\kms$ is the sound speed. The value of $\Tax(R_c)$ is set by hydrostatic equilibrium constraints at $r>R_c$ to be comparable to $T_{\rm vir}$, and is thus not truly a free parameter (see below). Eqn.~(\ref{eq:isentropic_sols}) shows that in this isentropic limit, temperature increases logarithmically inwards, while density and velocity increase somewhat faster than logarithmically. Thus, at inner radii, densities are lower and temperatures are higher than in the cooling flow limit derived above.

\begin{figure*}
\includegraphics[width=\linewidth]{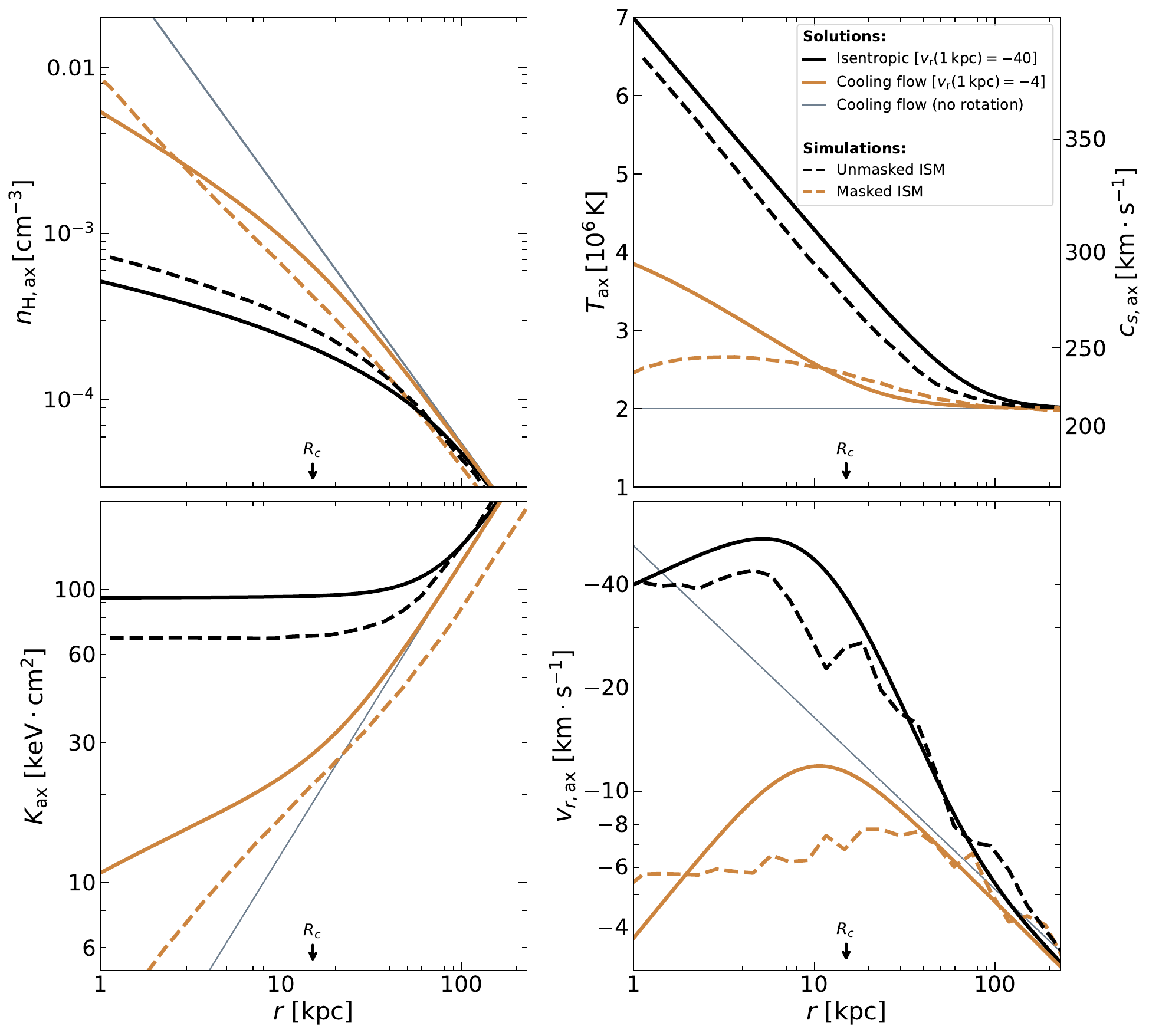}
    \caption{Dependence of hot CGM properties on conditions at the disk-halo interface. Panels show the hydrogen number density, temperature, entropy and radial velocity profiles along the axis of rotation. Solid curves correspond to integration of the flow equations~(\ref{eq:mass_naa})--(\ref{eq:entropy_naa}) for two different assumed inflow velocities at $1\kpc$, yielding either an isentropic solution (black) or a cooling flow solution (gold). Assumed circularization radius  $R_c=15\kpc$ is indicated by arrows, and grey lines plot a non-rotating cooling flow solution for reference. The two solutions differ by an order an magnitude in inner CGM density, despite having similar densities in the outer CGM.
    Dashed curves plot the results of the numerical simulations shown in Fig.~\ref{fig:gizmo_slices} below. 
    In the unmasked ISM simulation (black), $\vrax(1\kpc)=-40\kms$ due to numerical mixing at the disc-halo interface, while in the masked ISM simulation the inner boundary is at $1\kpc$ and $\vrax$ at this radius is set explicitly to $-5\kms$. The unmasked ISM simulation roughly realizes the isentropic solution while the masked ISM simulation roughly realizes the cooling flow solution.}
    \label{fig:analytic_vs_simulations}
\end{figure*}

\subsection{Numerical integration}\label{sec:analytic_results}
For the numerical integration of eqns.~\eqref{eq:mass_naa} -- \eqref{eq:entropy_naa}, we assume an isothermal potential with $v_c\!=\!200\kms$, a circularization radius of $R_c\!=\!15\kpc$, a uniform metallicity $Z\!=\!0.3\,{\rm Z_\odot}$ appropriate for the Milky Way CGM \citep{Miller15,Bregman18}, and $F_0 = 1$ in eqn.~\eqref{eq:fiducia_ftheta}.

We set the outer boundary at $\rout\!=\!300\kpc$, comparable to the virial radius of the halo, and set $\nhax(r_{\rm out})=1.1\times 10^{-5}\,{\rm cm^{-3}}$ and $\Tax(r_{\rm out}) = 2\times 10^6\K$, which is the temperature of a non-rotating cooling flow with the same $v_c = 200\kms$.
We note though that hydrostatic equilibrium implies the solution at $r\ll \rout$ is not sensitive to this choice \citep[see discussion in Appendix B of][]{Stern19}. The density at the outer boundary is chosen to be consistent with observational constraints on the Milky Way outer
CGM (see Section~\ref{sec:observations} below), and also implies an accretion rate of $\dot{M}=1.2\Msunyr$ in the cooling flow limit, comparable to the Milky-Way star formation rate of $\sim\!2\Msunyr$ \citep{BlandHawthornGerhard} assuming half of the mass of recently formed stars is ejected back to the ISM \citep{KrumholzDekel12}. 
We also choose $\rin=1\kpc$ as a characteristic size of the disk-halo interface. Under these conditions the solution is set by the chosen inflow velocity at the inner boundary $\vent$, which we vary below.

For the integration, we rewrite epns.~\eqref{eq:momentum_naa} and \eqref{eq:entropy_naa} as
\begin{equation}\label{eq:temperature_naa}
    \regd{\ln \Tax}{\ln r} = \frac{\tfz}{\gamma\tcz} - (\gamma-1)\frac{v_c^2}{\csax^2},
\end{equation}
\begin{equation}\label{eq:density_naa}
    \regd{\ln\rhoax}{\ln r} = -\frac{\tfz}{\gamma\tcz} - \frac{v_c^2}{\csax^2}.
\end{equation}
which together with eqn.~(\ref{eq:mass_naa}) form a closed set of differential equations. We employ the \texttt{solve\_bvp} routine from \texttt{SciPy} \citep{Kierzenka2001ABS,2020SciPy-NMeth}, which utilizes a fourth-order collocation algorithm that recasts the differential equations into a global root-finding problem, approximating the solution across the spatial domain with piecewise cubic splines, subject to boundary conditions. The algorithm iteratively subdivides the domain, adding grid points in regions of high numerical error, until the continuous residual falls below a specified tolerance. We verified that retaining the inertial terms in eqn.~\eqref{eq:momentum_r_axisymmetry}, proportional to $\mathcal{M}^2 = v_r^2/c_s^2\ll1$, makes a negligible change to the solution.

Figure~\ref{fig:analytic_vs_simulations} shows solutions for two values of $\vent$, namely $\vent=3\rin/\tcz(\rin)\!\approx\!4\kms$ (`cooling flow', gold) and $\vent=40\kms$ (`isentropic solution', black).  
In the low $\vent$ solution, the density (top left panel) follows the non-rotating cooling flow solution at $r\!\gg\!R_c$ (grey lines) and continues to rise inwards at $r<R_c$, albeit with a shallower slope than the non-rotating case, reaching $\sim\!5\times10^{-3}\,{\rm cm^{-3}}$ at $r=1\kpc$. The shallower slope is a result of the deflection of mass from the axis by rotation, and is comparable to the analytic approximation in this case (eqn.~\ref{eq:tf_over_tc}). The temperature rises modestly from the $2\times10^6\K$ at $r>R_c$ to $\sim\!4\times10^6\K$ at $r=1\kpc$ (top right panel), indicating that compressive heating overcomes radiative cooling. Because the inner temperature is higher and the inner density is lower than the non-rotating cooling flow, the entropy of the low $\vrax(\rin)$ solution is higher in the center, decreasing inwards with a slope of only $\tfz/\tcz\approx 1/3$ (eqn.~\ref{eq:entropy_naa}; bottom left panel), shallower than unity expected in a non-rotating cooling flow. The radial velocity follows the non-rotating solution at $r>R_c$, reaching $\vrax\!\sim\!-10\kms$ before decelerating towards the disk, approaching $\vrax\!\sim\!-4\kms$ at $r=1\kpc$, as expected from the analytic solution in eqn.~(\ref{eq:tf_over_tc}). 

The flow properties in the high $\vent$ solution show qualitative differences from the cooling flow solution out to $r\!\sim\!70\kpc$, significantly beyond $R_c$. The density increases only weakly with decreasing radius, reaching $5\times10^{-4}\,{\rm cm^{-3}}$ at $r=1\kpc$, ten times smaller than the density in the low $\vrax(\rin)$ solution. The temperature profile shows a logarithmic increase, reaching $\sim\!7\times10^6\K$ at $r=1\kpc$, as suggested by eqn.~\eqref{eq:isentropic_sols}. 
The entropy profile of this solution is flat, hence the name, since hot gas flows inward on a timescale much shorter than its cooling timescale. The entropy flattens already at $\sim60\kpc$, 
illustrating how the inner boundary condition is communicated to large radii in this subsonic flow. 

The mass inflow rate $\dot{M}$ in the two solutions is roughly the same and equal to the cooling flow rate with $\dot{M}\approx 1\Msunyr$, since the factor of $\sim\!10$ higher velocity in the isentropic solution is compensated by a factor of $\sim\!10$ reduction in  density (top left panel of Fig.~\ref{fig:analytic_vs_simulations}). A sink with a sufficiently large imposed velocity would however cause an increase of $\dot{M}$ beyond the cooling flow value, as the mass demand at the sink would exceed that supplied by cooling.

In summary, Fig.~\ref{fig:analytic_vs_simulations} demonstrates how the flow near the axis is highly sensitive to the physical conditions at the inner boundary. A value of $\vrax(\rin)$ consistent with a cooling flow produces a large density in the inner CGM and an inward-decreasing entropy profile, while a higher $\vrax(\rin)$ produces a factor of $\sim10$ lower density in the inner CGM and a flat entropy profile. 

\section{Idealized simulations}\label{sec:idealized_simulations}

\begin{figure*}
    \moveleft0.5cm\hbox{\includegraphics[width=1.05\linewidth]{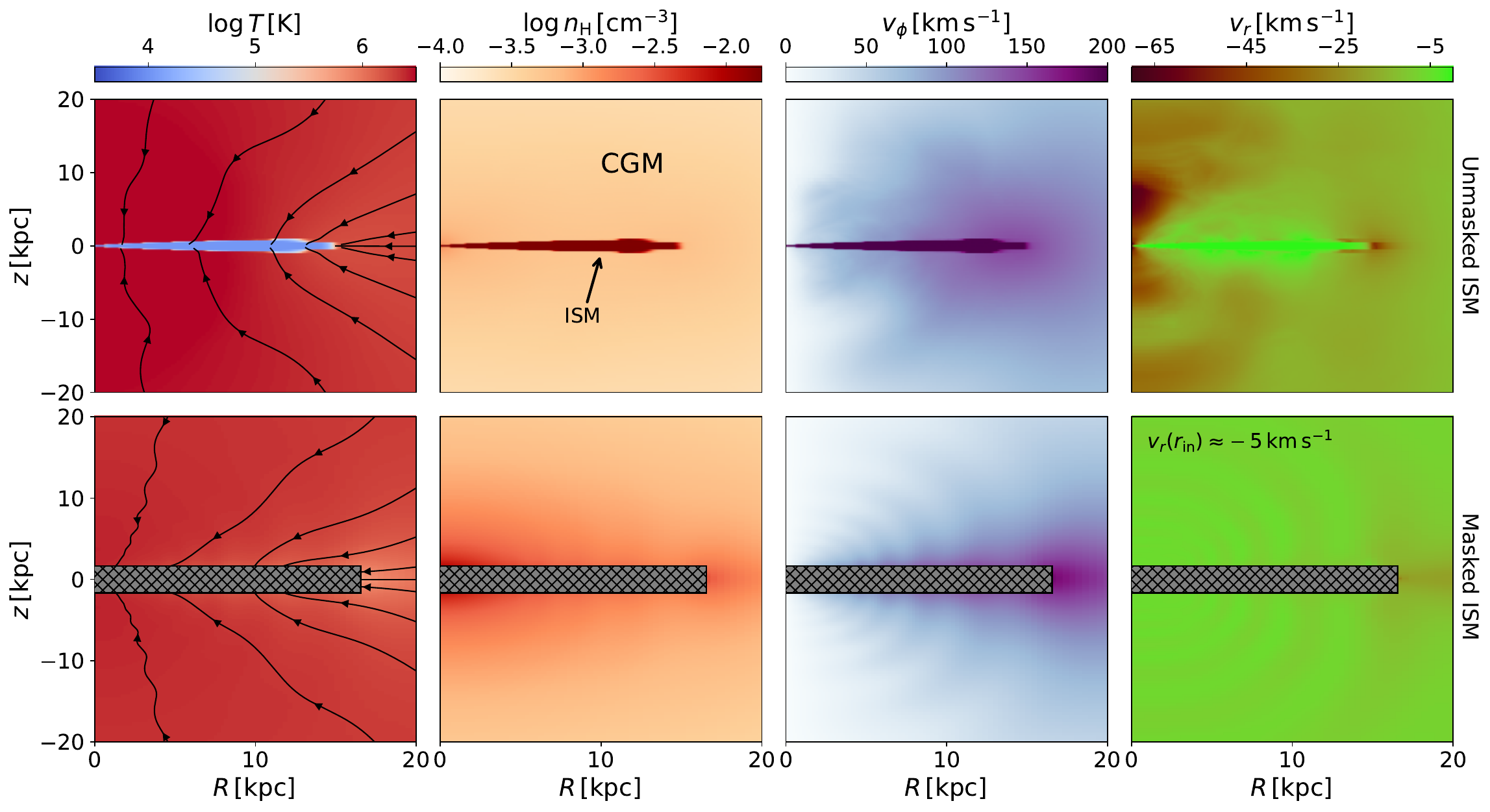}}
    \hspace*{-100cm}
    \caption{Two simulation setups for modeling hot CGM. Top row shows a simulation in which the cool ISM disk is within the simulated domain (`unmasked ISM'), similar to cosmological simulations and many idealized CGM simulations. The bottom row shows a simulation with an inner boundary outside the ISM (`masked ISM'), with the region not simulated marked with a hatched box. A slow inflow with $v_r=-5\kms$ is allowed through this inner boundary, corresponding to the hot CGM cooling time. From left to right, the panels plot  gas temperature, density, rotation velocity and radial velocity in the $R-z$ plane, averaged over the $\phi$ coordinate. Black curves in the left panels depict flowlines.
    The unmasked ISM simulation exhibits $\sim\times10$ higher inflow velocities and $\sim\times10$ lower densities  above the center of the disk, relative to the masked ISM simulation, due to numerical mixing at the interface.}
    \label{fig:gizmo_slices}
\end{figure*}
In this section, we present results from idealized 3D hydrodynamic simulations of hot and rotating CGM, in which the cool ISM disk forms a sink for hot gas. We run two types of simulations. In the first type the cool ISM disk is included in the simulated domain, similar to cosmological simulations and to many idealized CGM simulation setups \citep{Barbani23,Kopenhafer23,Stern24,Ranita24}. Below we show that in these simulations the disk-halo interface is unresolved at the rotation axis, and the hot-to-cool mass flux there is set by numerics and is resolution-dependent. We call this type \textit{unmasked ISM} simulations, and they are described in
Section~\ref{sec:cold_ism_sim}. In the second type of simulations, 
the region that would have formed a disk is replaced by an appropriate inner boundary condition, so the mass flux can be set to any desired value. These are similar in spirit to \cite{Fielding17} and \cite{Lochhaas20}, (though they did not include rotation) and are called \textit{masked ISM} simulations. They are described in Section~\ref{sec:masked_disk}.

\subsection{Unmasked ISM simulation}\label{sec:cold_ism_sim}
Our unmasked ISM simulations roughly follow the idealized setup in \citet{Stern24}. We 
use the Eulerian hydrodynamic code \texttt{Athena++} \citep{Stone20}, solving the hydrodynamics equations assuming an ideal gas with adiabatic index $\gamma = 5/3$. Included source terms are an analytic isothermal potential and optically thin cooling. For the cooling we use the \citet{Wiersma09} tables, which include also photoionization heating by a $z=0$ UV background as done above for the numerical integration. We initialize each simulation with density, temperature and radial velocity set by the non rotating cooling flow solution given by eqn.~\eqref{eq:spherical_solutions} with $\vc=200\kms$, $\dot{M}=1\Msunyr$ and uniform $Z=0.3\,{\rm Z_\odot}$, as in Section \ref{sec:analytic_results}. On top of this solution we add rotation with an angular momentum profile that depends on $\theta$ but is independent of $r$, namely $j = v_c R_c\sin^2\theta$, resulting in a $\theta$-independent angular velocity $\Omega = v_cR_c/r^2$ for $r\!>\!R_c$. For $r\!<\!R_c$, we set a constant $\Omega = v_c/R_c$. Here and in Section~\ref{sec:masked_disk} below, the exact choice of boundary conditions at $r\!\gg\!100\kpc$ does not affect the solution at $r\ll100\kpc$, as this region is well beyond the cooling radius, and therefore does not participate in the flow for the duration of the simulations \citep{Bertschinger89,Stern19}.

Our simulation box spans $|x|,|y|,|z|<500\kpc$ with a base grid size of $64^3$, refined statically in two nested cubes. A cube of side length $\!\sim\!30\kpc$ is refined to level 4 above the base grid, giving a cell size $\Delta x\sim\!480\,{\rm pc}$, and an inner cube of side length $\sim\!8\kpc$ is refined further to level 7, giving $\Delta x\sim\!60\,{\rm pc}$ at the cold-hot interface. We run our simulation for $\sim 10\,{\rm Gyr}$, which assures the simulation has reached a steady state out to $100\kpc$. The region within $3R_c=45\kpc$ reaches a steady state already after $t = \tcool(r=3R_c)\sim3.5\,{\rm Gyr}$ (eqn.~\ref{eq:spherical_solutions}; \citealt{Stern19}). The simulation adjusts to $\dot{M}=1.2\Msunyr$, slightly larger than the $1\Msunyr$ in the initial non-rotating solution. 

The top row of Figure~\ref{fig:gizmo_slices} shows, from left to right, the temperature, hydrogen number density, azimuthal and radial velocities in the meridional plane, averaged between $9\,{\rm Gyr}<t<10\,{\rm Gyr}$ and over $\phi$. Notice the large radial velocities near the axis before the gas joins the center of the disk, peaking at $|v_r|\!\sim\!60\kms$, and the low densities of $\nh\!<\!10^{-3}\,{\rm cm^{-3}}$ throughout the plotted domain. The temperature and density of the inner CGM in the unmasked ISM simulation are comparable to those in \citet[][Fig.~4]{Stern24}, while the velocities along the axis are slightly lower than their $|\vrax|\!\sim\!100\kms$. Below we show this is a result of resolution, where lower spatial resolution leads to higher velocities. 

In Fig.~\ref{fig:analytic_vs_simulations}, dashed black curves show the flow quantities near the axis in the simulation, calculated from cells with $\theta\!<\!20^\circ$ or $\theta\!>\!160^\circ$. The density (top left panel) increases only slowly, from $4\times10^{-5}\,{\rm cm^{-3}}$ at $r\!\sim\!100\kpc$ to $7\times10^{-4}\,{\rm cm^{-3}}$ at $r=1\kpc$. The temperature rises logarithmically from the virial temperature to $\sim\!6.5\times10^6\K$, characteristic of an isentropic flow (eqn.~\ref{eq:isentropic_sols}), which is confirmed by the flat entropy profile in the bottom left panel. The radial velocity (bottom right panel) increases steeply to $\sim\!40\kms$ near $R_c$, and keeps this value down to $r=1\kpc$. 

The unmasked ISM simulation is in good agreement with the $\vrax(\rin)=-40\kms$ isentropic flow solution (black solid), with $\sim\!15\%$ deviation in density and $\sim\!20\%$ in the velocity within $R_c$. Deriving the isentropic solution  with $F_0 =0.9$ rather than $F_0 =1$ generates an even better agreement, but the qualitative results are the same. In particular, the gas density at $r\sim1\kpc$ is  $\nhax\!\sim\!7\times10^{-4}\,{\rm cm^{-3}}$ with $\Tax\!\sim\!6.5\times10^6\K$ at $r\!\sim\!1\kpc$, implying a long cooling time of $\tcz\!\sim\!12\,{\rm Gyr}$. Nevertheless, this dilute hot gas is able to cool down to $10^4\K$ as it joins the disk within a $10^4$ times shorter timescale of $\Delta x/v_r=1.5\,{\rm Myr}$.

\begin{figure}
    \centering
    \includegraphics[width=1\linewidth]{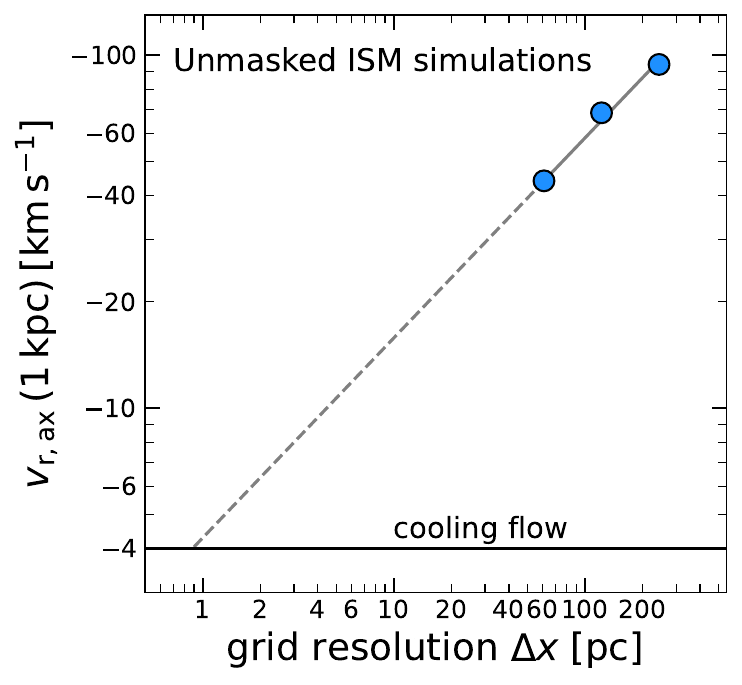}
    \caption{Hot CGM velocity versus grid resolution in the unmasked ISM simulations. Inflow velocity is measured at the axis at $r=\!1\kpc$. The line shows the best-fitting power law, $\vrax\propto\Delta x^{0.56}$, and its extrapolation down to the cooling flow velocity of $4\kms$, which is reached at $\Delta x=1\,{\rm pc}$. The high inflow velocity of the hot CGM in these simulations, and the resulting isentropic profile, are thus likely not physical but rather an artifact of limited resolution. }
    \label{fig:vr_vs_dx}
\end{figure}

\subsection{Numerical mixing at the cooling layer}\label{sec:numerical_mixing}
How does the hot gas cool so rapidly relative to its radiative cooling rate? 
The long radiative cooling time suggests that numerical mixing between the hot and cold gas allows the gas to cool. Hot gas flows directly into the cold disk before cooling, and the cold gas acts as an effective sink for the hot gas, by rapidly radiating the thermal energy it receives. Since numerical mixing depends on resolution, we expect the sink properties and thus the inner velocity to change with changing grid size. 

Figure~\ref{fig:vr_vs_dx} shows the radial velocity at $r=1\kpc$ along the axis in unmasked ISM simulations run with the same refinement structure as above, with three different maximum resolutions, namely $60$, $120$ and $240\,{\rm pc}$ (circles). The velocity scales approximately as $\Delta x^{0.56}$, decreasing with improving resolution, consistent with a flow driven by numerical mixing \citep{Tan21,Lancaster26}. Extrapolating this trend down to the cooling flow velocity of $4\kms$ implies a required resolution of $\Delta x\approx\!1\,{\rm pc}$ in order to realize a cooling flow solution in this setup.

We conclude that an unresolved, $\sim\!1\,{\rm pc}$-thick cooling layer at the disk-halo interface 
induces numerical mixing between the hot and cool gas. This in turn drives an isentropic dilute flow near the axis, which is more dilute than a cooling flow. 
We find a similar behavior in simulations with the same physical setup run with the Eulerian code \texttt{RAMSES} \citep{Teyssier02} and with the Lagrangian code \texttt{GIZMO} \citep{Hopkins15}.

\subsection{Masked ISM simulation}\label{sec:masked_disk}
While our unmasked ISM simulation is unable to realize the cooling flow solution due to the high resolution required as shown in Fig.~\ref{fig:vr_vs_dx}, replacing the disk-halo interface with an inner boundary condition for the CGM allows us to recover it. 

We run an \texttt{Athena++} simulation with a similar physical setup to the unmasked ISM simulation above, but using cylindrical coordinates. This allows us to exclude the region of the disk from the computation domain, and impose inner boundary conditions. To this end, we mask out the region of the disk, namely a cylinder spanning $0\!<\!R\!<\!17\kpc$ and $0\!<\!z\!<\!1\kpc$. The dimensions of the cylinder are chosen to prevent the formation of the cool structure which appears in the unmasked simulations (see Fig.~\ref{fig:gizmo_slices}). We treat the outermost two layers of the masked box as an effective ghost zone, similar to ghost zones at the edges of the simulation box. We perform quadratic extrapolation of all of the flow variables from the active cells into these ghost cells, except for the vertical velocity at the upper cap of the box, on which we impose $v_z = -(1\kpc)/\tc$, which is the flow time expected in the cooling flow solution. The remaining cells in the masked box are set to zero values for the velocities, and the density and pressure are chosen to result in a minuscule speed of sound, so as not to affect the time step. For the remaining boundary conditions, we use periodic boundary conditions in the azimuth, reflecting boundary conditions at $R=0$ and $z=0$, and at the outer boundaries fix the spherically radial and azimuthal velocities, temperature and density to their initial values.
The simulated domain spans $0<R<500\kpc$ and $0<z<500\kpc$, using 256 geometrically spaced cells in each direction and four cells across $2\pi$ radians in the azimuthal direction. The bottom row of Fig. \ref{fig:gizmo_slices} shows the temperature, hydrogen number density, azimuthal and radial velocities in the meridional plane of this setup, after the simulation has reached a steady state. For illustration purposes, the meridional plane is mirrored about the mid-plane. Note that the flow along the axis is slower and denser than in the unmasked ISM simulation shown in the top row, with roughly an order of magnitude difference.

Fig.~\ref{fig:analytic_vs_simulations} plots properties of the flow  near the axis in the masked ISM simulation. 
When computing these profiles, we smooth over a standing wave generated by reflections off the inner boundary, which amplitude is a few percent (most visible in the radial velocity panel of Fig.~\ref{fig:gizmo_slices}). The flow has the characteristics of a cooling flow solution (solid gold curves), with large densities, inward-declining entropy and small velocities. 
The simulation agrees qualitatively across all quantities with the cooling flow solution,  with deviations of a factor of $\lesssim2$ mostly in the velocity. We note that numerically integrating the 1D solution with $F_0=0.5$ improves the agreement with the simulation, but the qualitative behavior is the same.

The masked ISM simulation thus recovers the cooling flow solution, demonstrating that this solution can be realized in a numerical simulation provided that the inflow velocity onto the ISM is sufficiently small.

\begin{figure*}
    \centering
    \includegraphics[width=1\linewidth]{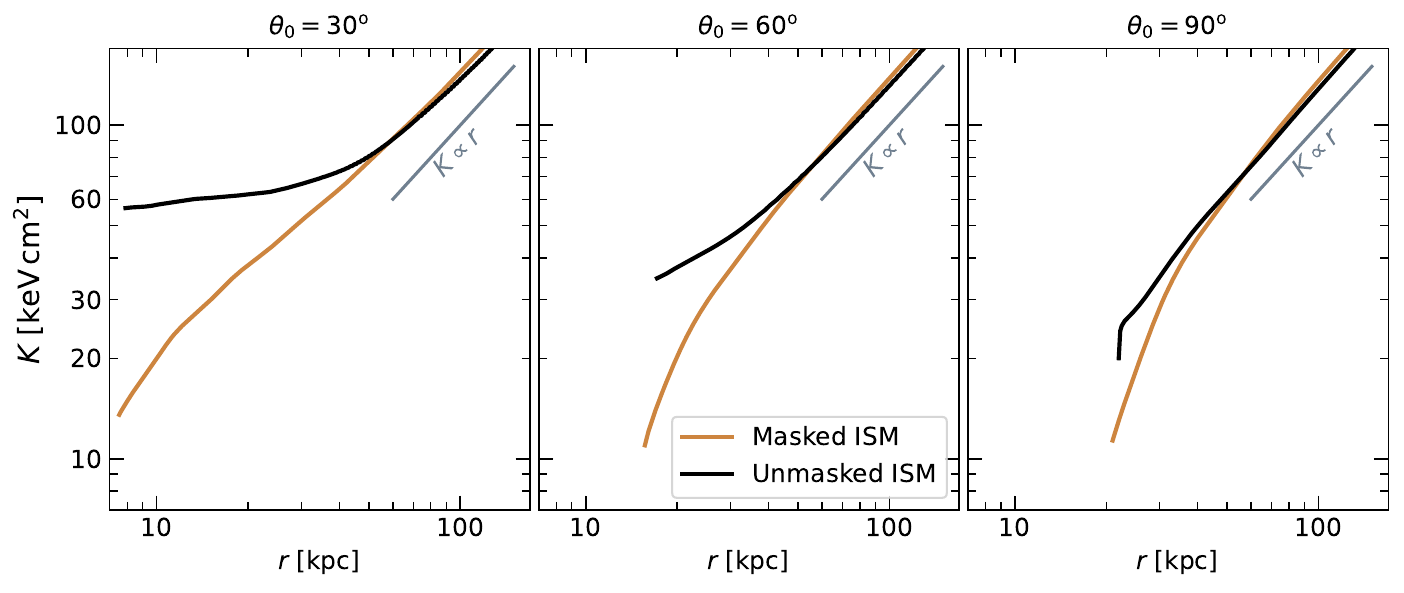}
    \caption{Sensitivity of hot CGM properties to conditions at the disk-halo interface, at off-axis angles. Entropy along flowlines originating at polar angles $\theta_0=30^{\circ},60^{\circ},90^{\circ}$ (left to right columns) is shown for both simulations. In the masked ISM simulation (gold) the entropy decreases inwards along all flowlines. In the unmasked ISM simulation (black) the entropy flattens within $r\lesssim40\kpc$ at $\theta_0=30^\circ$, but flattens less at $\theta_0=60^\circ,90^\circ$. Hot CGM properties are thus most sensitive to disk-halo conditions near the axis.}
    \label{fig:flowline_figure}
\end{figure*}

\subsection{Off-axis sensitivity to the conditions at the disk-halo interface}\label{sec:flowline_analysis}
Above we demonstrated the high sensitivity of hot CGM structure near the axis to the inflow velocity at the disk-halo interface. We now examine how off-axis regions of the hot CGM depend on the inner inflow velocity, by following flowlines from large radii down to the mid-plane in both the unmasked and masked ISM simulations. Flowlines are obtained by integrating the meridional velocity field $(v_r,v_\theta)$ from an initial outer point $(r_0,\theta_0)$, adopting $r_0=150\kpc$. The black lines in Fig.~\ref{fig:gizmo_slices} show examples of flowlines in each of the two simulations.

Fig.~\ref{fig:flowline_figure} shows the entropy along flowlines with initial angles $\theta_0=30^\circ,60^\circ,90^\circ$, similar to the $\theta_0=0^\circ$ flowline shown in Fig.~\ref{fig:analytic_vs_simulations}. At large radii, where the flow is approximately spherical, the flowlines in both simulations follow a $K\!\propto\!r$ entropy profile, as expected in a spherical cooling flow (Appendix~\ref{sec:spherical_solutions}). Closer in, as the flowlines approach their circularization radius $R_c\sin^2\theta_0$ (derived from eqn.~\eqref{eq:am_profile} and angular momentum conservation), the behavior diverges and differs qualitatively between different $\theta_0$. For the flowline with $\theta_0=30^\circ$ (left column), the entropy in the unmasked ISM simulation transitions to an isentropic profile within $r\lesssim40\kpc$, whereas in the masked ISM simulation it continues to decrease towards the disk, similar to the entropy profiles along the rotation axis (see Fig.~\ref{fig:analytic_vs_simulations}). In contrast, the flowline with $\theta_0=90^\circ$, which traces the mid-plane, shows entropy decreasing towards the disk in both simulations. At the intermediate angle $\theta_0=60^\circ$ (middle column), the entropy in the unmasked ISM simulation decreases more slowly than in a cooling flow, while in the masked ISM simulation it continues to decrease with a slope of a cooling flow.

Although the entropy profiles in the unmasked ISM simulation differ qualitatively between flowlines, they all reach a similar inner velocity of $|v_s|\approx20-40\kms$. Near the axis ($\theta_0=0,30^\circ$), such a velocity produces an isentropic profile whereas away from the axis the same velocity produces a steeper entropy decline ($\theta_0=60^\circ,90^\circ$). The sensitivity of the entropy profile to the inflow velocity near the disk therefore depends on the location where a given flowline approaches the mid-plane.

The varying sensitivity of the entropy profiles to conditions at the inner boundary is likely a result of flow geometry, specifically by how the cross-sectional area $\delta A$ of a flowline changes inward. Since mass conservation gives $\rho v\,\delta A\equiv{constant}$ along a flowline, the mass flux $\rho v$ is $\propto\delta A^{-1}$. At large radii, where the flow is approximately spherical, $\delta A\propto r^{2}$ and thus $\rho v\propto r^{-2}$. At smaller radii, flowlines closer to the axis are deflected by centrifugal forces such that $\delta A\propto r^{p}$ with $p<2$, so $\rho v$ grows more slowly inwards (see left panels of Fig.~\ref{fig:gizmo_slices}). A large velocity imposed at the disk-halo interface must then be compensated by a correspondingly small density, which lengthens the cooling time and produces the isentropic flow found in Section~\ref{sec:analytics}. Flowlines farther away from the axis however have $\delta A\propto r^{p}$ with $p>2$, since they tend to converge near the mid-plane, so that $\rho v_s\propto r^{-p}$ increases steeply inwards. Consequently, a similarly large velocity at the inner boundary does not significantly suppress the density, and the cooling time remains comparable to the flow time. 



\section{FIRE simulations}\label{sec:fire_sims}
Cosmological simulations of Milky Way-like galaxies reach typical hot CGM resolutions
far coarser than the $\sim 1\,{\rm pc}$ required to resolve the cooling layer (Fig.~\ref{fig:vr_vs_dx}). We therefore expect a large hot inflow velocity and isentropic conditions  near the axis in such simulations, rather than the cooling flow solution. Here we show that galaxies in the FIRE simulations indeed exhibit such a flow near the axis.

The FIRE-2 simulations \citep{Hopkins18} were run with the \texttt{GIZMO} code \citep{Hopkins15}, using a meshless finite mass hydrodynamics method. We analyze the Milky Way-like galaxies \texttt{m12b,i,c,m} at $z=0$, as well as the ELVIS galaxies \citep{GarrisonKimmel}. These galaxies were chosen for the presence of a quasi-hydrstatic volume-filling hot phase in the CGM, and the presence of a stable, rotationally-supported cold disk \citep{Ji20,Stern21,Yu21, Sijie22, Hafen22,Gurvich23,Sultan25,Myrtaj26}.  With a typical mass resolution of $m_b=7100\,{\rm M_\odot}$, the effective spatial resolution is $(m_b/\rho)^{1/3}\!\approx\!600\,{\rm pc}$ for $\nh=10^{-3}\,{\rm cm^{-3}}$ appropriate for the inner hot CGM, much larger than the required parsec-scale mentioned above. In these simulations \citet{Sultan25} found that the spherically-averaged temperature and density profiles beyond $20\,\kpc$ are well described by a non-rotating cooling flow. 

We analyze snapshots up to a lookback time of $1\,{\rm Gyr}$. In each snapshot, we calculate the median flow properties of particles with $\theta\!<\!20^\circ$ or $\theta\!>\!160^\circ$ about the axis of rotation, defined to be the direction of angular momentum of the galaxy. The angular momentum of the inner CGM is aligned with the disk angular momentum to within $\approx\!20^\circ$. We bin the particles along the radial direction with $\Delta\log{r/R_{\rm vir}} = 0.1$. We note that the CGM in FIRE is multiphase, composed of gas with different temperatures. Since we are interested in the hot phase of the CGM, we use the same method as \citet{Sultan25} to filter out the cold phase. In short, the hot phase density and temperature modes in each bin are measured from the distributions of all particles with $T>10^5\K$ within that bin. Then, a particle belongs to the hot phase if its temperature and density are at most $0.5\,{\rm dex}$ away from the mode. The final profiles are calculated as the median of all hot phase particles in each bin. We found bins with $r\!<\!10\kpc$ to have less than ten particles, and therefore perform our analysis down to $r=10\kpc$.

\begin{figure}
    \centering
    \includegraphics[width=1.08\linewidth]{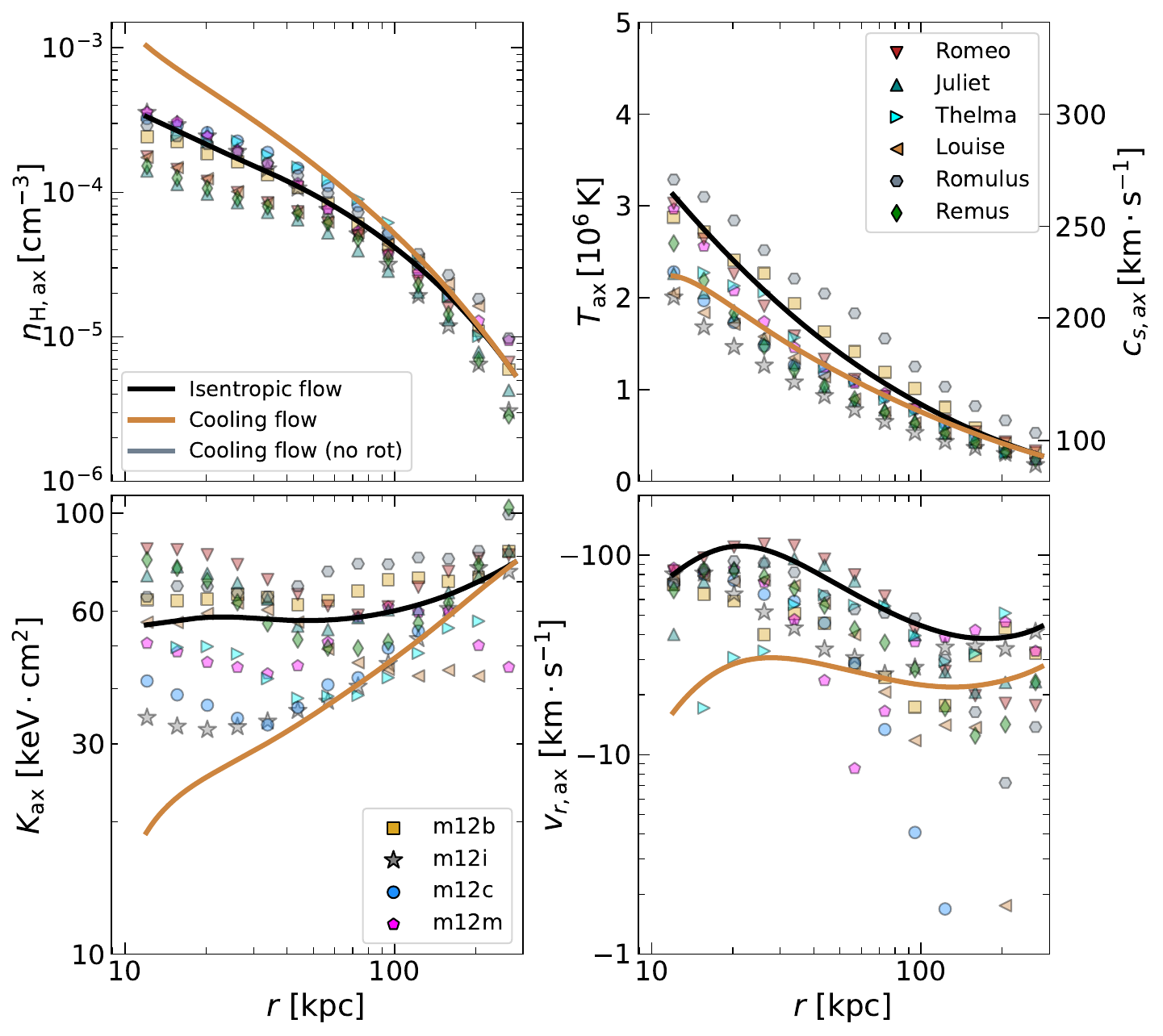}
    \caption{
    Comparison of FIRE simulations to the isentropic and cooling flow solutions. Symbols correspond to the measured quantities from different FIRE galaxies (see legend), including only hot phase particles near the axis of rotation. Solid curves correspond to the isentropic (black) and cooling flow (gold) solutions, with $\vrax(\rin)=-80\kms$ and $\vrax(\rin)=-16\kms$, respectively, for $\rin=10\kpc$. 
    Simulated hot CGM in FIRE follow the isentropic solution along the axis. A cooling flow cannot form along the axis in these simulations due to the required resolution.}
    \label{fig:analytic_vs_fire}
\end{figure}

Figure~\ref{fig:analytic_vs_fire} shows the hydrogen number density, temperature, entropy and radial velocity profiles along the axis of all FIRE galaxies analyzed, indicated by the markers. The densities of the FIRE galaxies increase with decreasing radius, with average densities increasing from $\!\sim\!5.5\times10^{-6}\,{\rm cm^{-3}}$ at $r=250\kpc$ to $\sim\!2\times10^{-4}\,{\rm cm^{-3}}$ at $r=10\kpc$ (top left panel). This increase, together with a temperature increase by a factor of $\sim\!10$, from $\sim\!2\times10^5$ at $r=250\kpc$ to $3\times10^6\K$ at $r=10\kpc$ (top right panel), results in a flat average entropy profile (bottom left panel). From eqn.~\eqref{eq:entropy_naa}, a flat entropy profile requires high velocities, consistent with the $\vrax\sim\!100\kms$ reached by the FIRE galaxies (bottom right panel), roughly eight times larger than expected from the non-rotating cooling flow fit to these galaxies \citep{Sultan25}.

Figure~\ref{fig:analytic_vs_fire} also shows the results of two numerical integrations of eqns.~\eqref{eq:mass_naa} -- \eqref{eq:entropy_naa} indicated by the solid curves. 
For calculating these solutions, we use the median circular velocity and metallicity profiles calculated from the simulations. These profiles, along with the turbulence profile described below, were computed in the same bins as the flow profile. We take into account turbulent pressure support, by swapping $\csax^2$ in eqn.~\eqref{eq:momentum_naa} with $\csax^2\lbrac{1+\alpha}$, where $\alpha = \gamma\sigma^2/\csax^2$, and $\sigma$ is the turbulent velocity dispersion. We find $\sigma\!\sim\!70\kms$ at $r<50\kpc$. \citet{Sultan25} found that including this non-thermal pressure support allows for a better fit to the non-rotating cooling flow of the spherically averaged profiles.

We set the outer boundary at $\rout=250\kpc$, which is roughly the cooling radius in these halos \citep{Sultan25}. The density and temperature are set to the average values of the FIRE galaxies at this radius, namely $\nhax(\rout)\!=\!5.5\times10^{-6}\,{\rm cm^{-3}}$ and $\Tax(\rout)\!=\!2.8\times10^5\K$. For the inner boundary condition at $\rin=10\kpc$, we set $\vrax(\rin)=-80\kms$ for the isentropic solution (black curves in Fig.~\ref{fig:analytic_vs_fire}), which is the average velocity of the FIRE galaxies. For the corresponding cooling flow solution (gold curves in Fig.~\ref{fig:analytic_vs_fire}), we set$\vrax(\rin)$ equal to $-\rin/\tcz(\rin)\approx -16\kms$.

The isentropic solution reproduces the average trends in the simulations over the full range of radii considered. In particular, the entropy profile is nearly flat, rising merely as  $\sim r^{0.06}$ at radii smaller than $100\kpc$. In contrast, the cooling flow solution produces a steeper density profile, reaching $\nhax(\rin)\!\sim\!10^{-3}\,{\rm cm^{-3}}$, five times the average density of the FIRE galaxies. The cooling flow solution also has an increasing entropy profile, rising as $\sim\!r^{0.4}$. This is shallower than $r^{0.8}$ produced by the non-rotating cooling flow solution of \citet{Sultan25}, due to the effects of rotation (see Section \ref{sec:analytic_results} above).

The large velocities and flat entropy profiles implies that the flow along the axis in FIRE galaxies is better described as an isentropic flow than a cooling flow. A potential origin for this is turbulent mixing, since the velocity dispersion of the gas in the ISM of these galaxies is $\sigma_g\!\sim\!30\kms$ \citep{Gurvich20}. If the turbulent mixing layer has a similar turbulent velocity, the expected entrainment velocity is $v_{\rm in}\approx 30-50\kms$ \citep{Fielding20}, a factor of $\sim\!10$ higher than the velocity required for a cooling flow. However, it is unclear whether the simulations capture this entrainment correctly, as the Field length of $\approx15$ pc is highly unresolved. It has been argued that resolving the largest eddy may be enough to properly capture the inflow onto the turbulent layer, due to a cancellation between numerical diffusivity ($\propto \Delta x^{1/2}$) and the numerical viscosity ($\propto \Delta x^{-1/2}$) which results in a resolution-independent energy flux onto the layer \citep{Tan21, Lancaster26}. In FIRE, even the largest eddy which is of the order of the disk scale height $\approx 500\,{\rm pc}$ \citep{Gurvich20}, is only marginally resolved by the hot gas which has $\Delta x=600\,{\rm pc}$. It is thus uncertain if the simulated flow is indeed set by turbulent mixing or by numerical limitations.



\section{The Milky Way hot CGM}\label{sec:mw_cgm}

Above we showed that the cooling flow and isentropic flow solutions differ by an order of magnitude in density for $r\!\lesssim\! R_c$, corresponding to a factor of $\sim\!100$ in emission measure, so observations near the rotation axis could in principle distinguish between them. X-ray observations above the disk are available for the Milky Way CGM, a region where the emission is dominated by the eROSITA bubbles \citep{Predehl20}. These bubbles are widely believed to be shock-heated gas driven by an outflow \citep[see][for a review]{Sarkar24}. Since our models do not include a shock and post-shock gas, we focus on comparing our models with the observationally-inferred Milky-Way entropy profile, since the effect of the shock on entropy is less model-dependent than on other quantities such as density.  
To this end, in Section~\ref{sec:observations} we compile density and temperature estimates along the axis from the literature, and infer the entropy. In Section~\ref{sec:mw_model}, we obtain the predicted entropy profiles for the isentropic and cooling flow solutions appropriate to the Milky Way, add the effect of a shock, and compare them to observations.

\subsection{Observational constraints on Milky-Way hot CGM entropy profile}\label{sec:observations}

The density of the hot CGM of the Milky Way has been constrained by a number of independent methods. Ram-pressure stripping models of Milky Way satellites are used to estimate the CGM density at the pericentric passages of these satellites, independently of the temperature and ionization state of the ambient gas. \citet{Grcevich} estimated pericentric densities for Carina, Ursa Minor, Sculptor and Fornax based on proper motions, probing mostly $r\!\sim\!50-100\kpc$. They estimated $\nh\!\approx\!0.4-1\times 10^{-4}\,{\rm cm^{-3}}$ at these radii, though their uncertainties span an order of magnitude in both location and density. Tighter constraints were obtained by \citet{Gatto13} and \citet{Salem15}, who combined analytic estimates with hydrodynamic simulations of the ram-pressure stripping of Carina, Sextans and the Large Magellanic Cloud, finding $n_{\rm H}\!\sim\!(4-7)\times10^{-5}\,{\rm cm^{-3}}$ at $r\!\sim\!50-80\kpc$.

Deep observations of soft X-ray emission have detected absorption and emission from O\,{\sc vii} and O\,{\sc viii}, which most likely originate in the hot gaseous halo of the Milky Way \citep{Yoshino09,Gupta12,Henley13,Miller13,Miller15,Li17}. Assuming optically thin O\,{\sc vii} absorption and a constant $n_{\rm OVII}/n_e$, \citet{Miller13} fit a cored $\beta$-model, $n = n_0\lbrac{1+(r/r_{\rm core})^2}^{-3\beta/2}$, and found $\beta = 0.56^{+0.10}_{-0.12}$, which for $r\gg r_{\rm core}\approx 2\kpc$ yields a power law consistent with the density slope of a non-rotating cooling flow \citep{Stern19}. Similar slopes were found also by later studies from the same group \citep{Miller15,Bregman18}. 

To constrain the densities at large radii we adopt the fit from \citet{Miller13}, using the corrected profile given in \citet{Voit19}. \citet{Voit19} renormalized the \citet{Miller13} profile to account for a temperature-dependent O\,{\sc vii} ionization fraction, adopting temperatures that rise from $\approx10^6\K$ at $r=100\kpc$ to $\approx3\times10^6\K$ at $r=1\,{\rm kpc}$, while remaining consistent with the constraints of \citet{Gupta12}. Because these fits assume spherical symmetry, they constrain only the angle-averaged density profile. We therefore use them to constrain the axial densities, and in turn, the entropy, only at large radii where our solution is spherical.

The observational studies just mentioned typically avoid the region with Galactic longitudes $|l|<60^\circ$ because of the Fermi/eROSITA bubbles \citep{Su10,Predehl20}. Density constraints close to the disk and rotation axis thus instead come from observations of the bubbles. 
\cite{Gupta23} (henceforth \citetalias{Gupta23}) analyzed the soft X-ray emission along a set of \textit{Suzaku} sightlines, deriving emission measures throughout the entire sky. We focus on emission measures from sightlines within Galactic latitude $|b|\leq45^\circ$ and longitude $|l|\leq60^\circ$. We focus on the emission measures of the $k_BT\approx 0.2\,{\rm keV}$ component (top row in their Supplemental~Fig.~2), which typically contributes $\approx\!80\%$ of the emission measure along each sightline. For each $|b|$ considered, we combine the northern and southern emission measures by averaging the two median values (red crosses in Supp.~Fig.~2 in \citetalias{Gupta23}). We define the uncertainty on the median value as the full range spanned by the lowest and highest values allowed by the error bars of the two measurements. This yields median $EM=0.02\pm0.006\,\emunits$ at $|b|\approx15^\circ$ and $EM=0.01\pm0.006\,\emunits$ at $|b|\approx45^\circ$. These latitudes correspond to distances of $r=2$ and $8\kpc$ along the axis, respectively. Adopting $Z=0.3\,{\rm Z_\odot}$ \citep{Miller15,Li17,Kataoka18,Martynenko} rather than the solar metallicity assumed by \citetalias{Gupta23} increases the inferred emission measures by a factor of three\footnote{Because the emission at $T\sim10^6\K$ is dominated by metal lines, $EM$ is approximately $\propto Z^{-1}$.}, giving $EM(|b|=15^\circ)=0.06\pm0.017\,\emunits$ and $EM(|b|=45^\circ)=0.03\pm0.02\,\emunits$. We assume $Z=0.3\,{\rm Z_\odot}$ also below.

The densities along the axis are derived from the emission measure using $\nh=\sqrt{EM/d}$, where $d$ is the assumed path length. We adopt $d=r$ for the sightlines, corresponding to $r=2\kpc$ and $8\kpc$, respectively. Similar conversions from emission measure to density were performed by \citet{Gupta23} and \citet{Sarkar24}. The uncertainties in the densities are estimated by including the measured emission-measure uncertainties and allowing the assumed path lengths to vary by a factor of two. This  gives $\log n(2\kpc)=-2.25\pm0.2$ and $\log n(8\kpc)=-2.67\pm0.3$.

\begin{figure}
    \centering
    \includegraphics[width=1\linewidth]{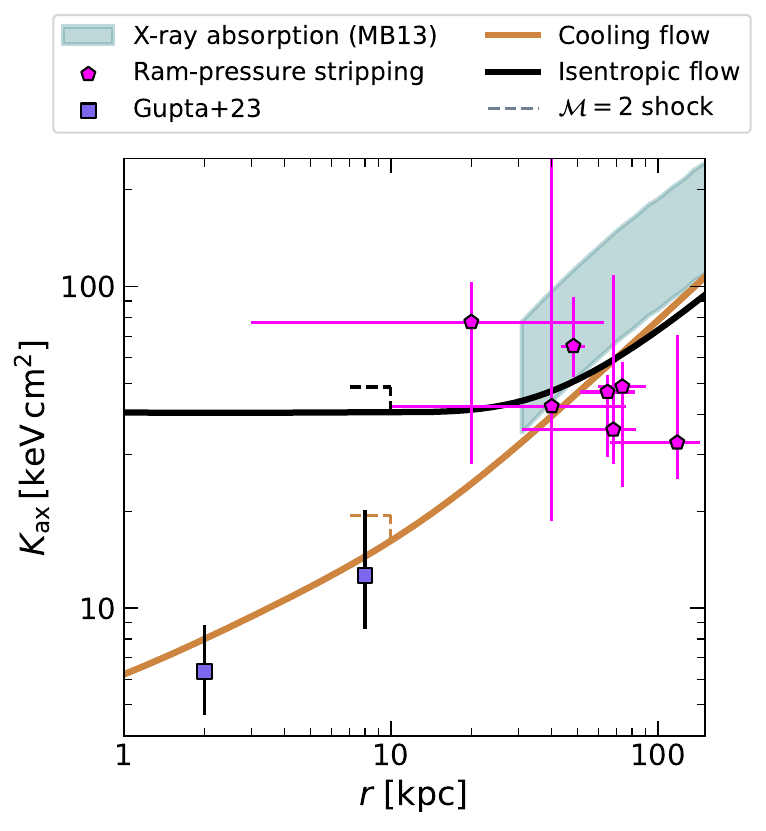}
    \caption{Contrasting the entropy profiles of the isentropic (black) and cooling flow (gold) solutions with observational constraints on the Milky Way hot CGM. 
    Dashed lines indicate the effect of including a $\mathcal{M}=2$ shock at $r=10\kpc$ as suggested by the eROSITA bubbles (see text). Square markers show the entropy inferred from the emission measures of \citet{Gupta23}. The shaded region is based on O\,{\sc vii} absorption measurements from \citet{Miller13}, and entagons show ram-pressure stripping-based estimates from \citet{Grcevich}, \citet{Gatto13} and \citet{Salem15}, assuming $T=10^6\K$. Observations imply a drop in entropy from large to small radii, similar to a cooling flow, and in tension with an isentropic flow. The presence of a shock makes this tension stronger.}
    \label{fig:observations}
\end{figure}

We convert the densities above to entropy assuming $T\approx 10^6\K$ at $r=70\kpc$ (see FIRE simulations above), and $k_BT=0.2\,{\rm keV}$ for the \citetalias{Gupta23} sightlines. Figure~\ref{fig:observations} shows the resulting entropy for the ram-pressure stripping (purple hexagons), X-ray absorption (shaded region), and \textit{Suzaku} (purple squares) estimates. Together, they imply that the entropy along the axis declines by a factor of $\approx10$ between $r=70\kpc$ and $r=2\kpc$. We note that a $40\%$ lower temperature at $r=70\kpc$, as suggested by the \citet{Faerman20} model, would reduce the implied drop.

\subsection{Flow solutions for the Milky Way}\label{sec:mw_model}

To determine which of our two solutions can describe the flow near the Milky Way rotation axis, we solve eqns.~\eqref{eq:mass_naa}--\eqref{eq:entropy_naa} using the Milky Way potential from \cite{Bovy15}. For the outer boundary conditions at $\rout=70\kpc$ we set $\nhax(\rout) = 5\!\times\!10^{-5}\,{\rm cm^{-3}}$ to match the observational constraints from X-ray absorption (see above), and set $\Tax(\rout)=10^6\K$. For the inner boundary condition at $\rin=1\kpc$, we set either $\vrax(\rin)=3\rin/\tcz \approx 10\kms$ which gives the cooling flow solution, or $\vrax(\rin)=100\kms$ which gives the isentropic solution. This isentropic solution velocity is larger than the $\approx\!40\kms$ used in Section~\ref{sec:analytic_results} due to the softer potential assumed at inner radii. The rest of the parameters are kept the same as in Section \ref{sec:analytic_results}. 

The solid curves in Fig.~\ref{fig:observations} show the entropy profiles of the cooling flow (gold) and isentropic flow (black) solutions. The entropy in the cooling flow solution declines inwards for the entire range of radii considered, as suggested by the observations, while the entropy of the isentropic flow solution is flat out to $r\approx50\kpc$. Given the drop in entropy from $r=70\kpc$ to $r=2\kpc$ implied by observations, the isentropic flow is in tension with observations, regardless of any shock. Interestingly, the cooling flow solution is in good agreement with the inward-decreasing entropy profile implied by observations even without a shock. 

To demonstrate how a shock would affect the two solutions, we assume a $\mathcal{M}=2$ shock at $r=10\kpc$, as typically derived for the bubbles \citep{LaRocca20,Yamamoto,Sarkar24}. Such a Mach number corresponds to a $\approx 20\%$ increase in entropy behind the shock. The entropy after the shock is determined by the adiabatic Rankine-Hugoniot jump conditions, as the post-shock densities and temperatures imply a cooling time of $\gtrsim1\,{\rm Gyr}$ \citep{Predehl20}, much longer than current age estimates of $\lesssim 50\,{\rm Myr}$ of the bubbles \citep{Guo12,Fox15,Predehl20,Sarkar24}. The dashed lines in Fig.~\ref{fig:observations} show the resulting post-shock entropy for the cooling and isentropic flow solutions. The exact post-shock entropy profile depends on the details of the outflow, so we show only the expected entropy right behind the shock. With the shock, the isentropic flow is in even stronger tension with the observed values since the shock increases its entropy. The post-shock entropy of the cooling flow solution is within the uncertainty of the entropy measurement at $r=8\kpc$.

Comparing the two solutions to these observational estimates, the isentropic flow solution is in strong tension with observations, while the cooling flow agrees with the inward-decreasing entropy profile trend seen in the observations, even in the presence of a shock.
\section{Summary}\label{sec:summary}

In this paper we examine how processes at the disk-halo interface shape the thermodynamics and observational signatures of the hot CGM in Milky Way-mass halos. Such processes remain unresolved even in state-of-the-art cosmological simulations, a limitation which can cause simulated CGM structure to be determined by numerics rather than by physical conditions at the interface.

We start by demonstrating analytically that the near-axis CGM structure is governed by how fast hot gas joins the cool disk. When the gas flows in on its cooling timescale, a \textit{cooling flow} with an inward-decreasing entropy profile forms. In contrast, when the gas flows in much faster than it cools, an \textit{isentropic flow} forms. Physically, such a rapid inflow may be driven by entrainment into turbulent mixing layers \citep{Fielding20}, possibly produced by disk--halo shear \citep{Lin25} or galactic fountains \citep[e.g.,][]{Fraternali17,Li23}. Confronting the two solutions with simulations, we find that the near-axis CGM is better described by the isentropic solution in both idealized and cosmological simulations, potentially due to unresolved cooling at the disk-halo interface.

Our main findings are summarized as follows:
\begin{enumerate}
\addtolength\itemsep{-1mm}
    \item Near the rotation axis, a cooling flow produces an entropy profile of $K\propto r^{1/3}$ in the inner CGM, while the isentropic flow produces a flat profile. In our idealized setup, the cooling flow is $\sim\!10$ times denser than the isentropic flow, which translates to a possible factor of $\sim\!100$ difference in emission measure (Fig.~\ref{fig:analytic_vs_simulations}; Section~\ref{sec:analytic_results}).
    
    \item In idealized simulations of hot CGM inflows which include the ISM in the simulated domain, the cool ISM forms a sink for the hot gas, driving the flow towards the isentropic solution near the rotation axis. In contrast, a cooling flow forms when the ISM is replaced by an inner boundary condition that imposes an inflow timescale comparable to the cooling time (Figs.~\ref{fig:analytic_vs_simulations}, \ref{fig:gizmo_slices}; Section~\ref{sec:idealized_simulations}).
    \item Numerical mixing at the unresolved disk-halo interface rapidly cools and pulls in the hot gas. This renders the inner velocity resolution-dependent ($|v_r|\!\propto\!\Delta x^{0.56}$).~Extrapolating this trend indicates parsec-scale resolution is required to resolve the cooling layer (Fig.~\ref{fig:vr_vs_dx}; Section~\ref{sec:numerical_mixing}).
    
    \item The sensitivity of hot CGM properties to inner boundary conditions is most pronouced near the rotation axis, while it is reduced at larger polar angles, closer to the edge of the disk (Fig.~\ref{fig:flowline_figure}; Section~\ref{sec:flowline_analysis}). This is a result of flowlines diverging from the axis and converging towards the mid-plane due to the centrifugal force.
    Since most of the mass flux in rotating cooling flows approaches the disk from its edge, a simulated hot CGM with an unmasked ISM can still form a cooling flow in the mass-flux averaged sense, as found by previous studies \citep{Hafen22,Stern24,Sultan25,Sultan26,Sankar26}.
    
    \item An isentropic inflow better describes the near-axis CGM of FIRE-2 Milky Way-like galaxies. It is not clear if this due to turbulence at the disk-halo interface in FIRE, or due to the  spatial resolution of $\sim\!600\,{\rm pc}$ which is insufficient for a cooling flow to form (Fig.~\ref{fig:analytic_vs_fire}; Section~\ref{sec:fire_sims}).
    
    \item Emission measure observations across the eROSITA bubbles \citep{Gupta23} together with constraints on hot CGM entropy at large radii, imply an inward-decreasing entropy profile along the Milky Way rotation axis. This decrease is consistent with the cooling flow solution, and in strong tension with the isentropic solution. The tension is strengthened once we include a shock between the inflow and outflowing bubbles (Fig.~\ref{fig:observations}; Section~\ref{sec:mw_cgm}).
\end{enumerate}

Which conditions at the disk-halo interface hold in real galaxies remains an open question: emission measures of the Milky Way favor an inward-decreasing entropy profile consistent with a cooling flow, whereas cosmological simulations exhibit a flat entropy profile. Settling this requires parsec-scale resolution across a large fraction of the disk-halo interface, a significant challenge for numerical simulations.

\bibliography{sample701}{}
\bibliographystyle{aasjournalv7}


\appendix
\section{Spherical cooling flow}\label{sec:spherical_solutions}
In this section, we briefly summarize the results of hot, non-rotating cooling flow solutions, and refer to \citet{Stern19} for a detailed analysis. 

When rotation is negligible, the system becomes spherically symmetric, and in steady state, eqns. \eqref{eq:mass_full}--\eqref{eq:entropy_full} reduce to
\begin{equation}
    4\pi r^2\rho v_r = -\dot{M},
\end{equation}
\begin{equation}\label{eq:momentum_spherical}
    \frac{1}{2}\regd{v_r^2}{r} = -\frac{1}{\rho}\regd{P}{r} - \frac{v_c^2}{r},
\end{equation}
\begin{equation}\label{eq:entropy_spherical}
    v_r\regd{\ln K}{r} = -\frac{1}{\tc},
\end{equation}
where $\dot{M}$ is the constant mass inflow rate and $v_c^2 = r\dd\Phi/\dd r$. Multiplying eqn.~\eqref{eq:entropy_spherical} by $t_{\rm ff} = \sqrt{2}r/v_c$, we get
\begin{equation}
    \frac{\sqrt{2}v_r}{v_c}\regd{\ln K}{\ln r} = -\frac{t_{\rm ff}}{\tc}.
    \label{eq:entropy_sphericalb}
\end{equation}
 It follows from eqn.~(\ref{eq:entropy_sphericalb}) that in the limit $\tc\!\gg\! t_{\rm ff}$, the system admits either a cooling flow solution with $-v_r\approx r/\tcool$ \citep[e.g.,][]{MathewsBregman78,Fabian84} or an isentropic solution with $-v_r\gg r/\tcool$. 
In the main text above we showed that which of these two solutions is chosen depends on the inner boundary conditions, and in the rotating case also on the flowline geometry. 

For the cooling flow solution, in which $v_r^2\!\ll \!v_c^2\!\approx\! c_s^2$ where the sound speed is $c_s^2 = \gamma^{-1}P/\rho$, one can neglect the inertial term in eqn.~\eqref{eq:momentum_spherical} and the system admits power law solutions. For $v_c\equiv const.$, \citet{Stern19} found
\begin{equation} \label{eq:spherical_solutions}
    \begin{split}
        c_s^2 & = \frac{10}{9}v_c^2, \\
        \Tsph & = 2\times 10^6\,v_{c,200}^2\K,\\
        n_{\rm H,sph} & = 0.8\times10^{-3}r_{10}^{-3/2}v_{c,200}\dot{M}_1^{1/2}\Lambda_{-22}^{-1/2}\,{\rm cm^{-3}},\\
        \vrsph & =-27\,r_{10}^{-1/2}v_{c,200}^{-1}\dot{M}_1^{1/2}\Lambda_{-22}^{1/2}\kms,\\
        \tcsph = t_{\rm flow,sph} & = 370\,r_{10}^{3/2}v_{c,200}\dot{M}_1^{-1/2}\Lambda_{-22}^{-1/2}\,{\rm Myr}.
    \end{split}
\end{equation}
Here, $r_{10}=r/10\kpc$, $v_{c,200}=v_c/200\kms$, $\dot{M}_1=\dot{M}/1\Msunyr$, and $\Lambda_{-22}=\Lambda/10^{-22}{\rm erg\,cm^{3}\,s^{-1}}$. Since $\tcsph=\tf$ in the cooling flow, the gas loses entropy as it flows in, and the resulting entropy profile declines towards the center as $K\propto r$. 

The solutions in eqn.~\eqref{eq:spherical_solutions} are valid until $v_r\!\approx\!c_s$, which occurs at the sonic radius $r_{\rm sonic}$. At $r<r_{\rm sonic}$ the gas loses thermal pressure support, cooling rapidly to $10^4\K$, and \cite{Goldner26} showed that instead the dominant support term becomes turbulent pressure. The sonic radius can be estimated from the subsonic solution by setting $v_r=c_s$ in eqn.~\eqref{eq:spherical_solutions}
\begin{equation}\label{eq:rsonic}
    r_{\rm sonic}\approx 0.17\,v_{c,200}^{-4}\dot{M}_1\Lambda_{-22}\kpc.
\end{equation}
Note that for the Milky Way, $r_{\rm sonic}$ is well within the galaxy, and therefore the assumption $v_r^2\!\ll\!c_s^2$ is valid throughout the Milky Way's CGM. In this paper we focus only on systems where $r_{\rm sonic}$ is smaller than the galaxy scale.

The cooling flow solution is a steady state solution only approximately. It evolves on the age of the system, or the time since the last heating event. Thus, within the cooling radius $r_{\rm cool}$ where the cooling time equals the age of the systems, steady state conditions apply. The flow through this radius is by definition not steady, and requires a special treatment. \citet{Bertschinger89} found a time dependent, self-similar solution for cooling flows, also for $r>r_{\rm cool}$. Their solution shows that the flow is well approximated by the steady state solution for $r\ll r_{\rm cool}$.
\section{Analytical solutions to the near-axis equations}\label{sec:analytic_solutions}
In this section, we derive analytical solutions for eqns.~\eqref{eq:temperature_naa}-\eqref{eq:density_naa} for the cooling flow and isentropic solutions. We note that eqn.~\eqref{eq:mass_naa} can in general be integrated analytically, which for the assumed $F_\theta$ in eqn.~\eqref{eq:fiducia_ftheta} gives $r^2\eta\rhoax\vrax\equiv const.$, where
\begin{equation}\label{eq:eta}
    \eta=\exp\lbrac{-2\int\limits_{\infty}^r\frac{F_\theta}{r}\dd r} = \lbrac{1+\frac{5}{9F_0}\lbrac{\frac{R_c}{r}}^2}^{F_0}.
\end{equation}
For $r\ll R_c$ and $F_0=1$, $\eta\approx 5/9(r/R_c)^{-2}$.
\subsection{Cooling flow solutions}\label{sec:power_law_solutions}
Equations \eqref{eq:temperature_naa}-\eqref{eq:density_naa} admit power law solutions for $r\ll R_c$, where $F_\theta\approx F_0= const.$. For these solutions, assuming $v_c= const.$, we get $\Tax= const.$, and therefore from eqn.~\eqref{eq:temperature_naa}. It then follows that $\tfz/\tcz$ is constant for these solutions, that is these solutions are valid for a cooling flow.

We assume $\rhoax\propto r^{-\beta}$, and since $\tfz\propto \tcz$, $\tfz =r/\vrax$ and $\tcz\propto 1/\rhoax$ for a constant $\Tax$, we find $\vrax\propto r^{1-\beta}$. Plugging $\vrax$ and $\rhoax$ into mass conservation, eqn.~\eqref{eq:mass_naa}, and assuming $F_\theta\approx F_0$, we get 
\begin{equation}\label{eq:power_law_density}
    \beta=\frac{3}{2}\lbrac{1-\frac{2}{3}F_0}.
\end{equation}
For our fiducial $F_0=1$, we get $\beta = 1/2$ and $\vrax\propto r^{1/2}$. From eqn.~\eqref{eq:temperature_naa}, we find $\tfz/\tcz = \gamma(\gamma-1)v_c^2/\csax^2$. Plugging $\rhoax\propto r^{-\beta}$ into eqn.~\eqref{eq:density_naa}, we get
\begin{equation}\label{eq:power_law_ratios}
     \frac{v_c^2}{\csax^2} = \frac{9}{10}\lbrac{1-\frac{2}{3}F_0},~~~\frac{\tfz}{\tcz}=1-\frac{2}{3}F_0.
\end{equation}
For $F_0=0$, we recover the spherical solutions in eqn.~\eqref{eq:spherical_solutions}. Equation \eqref{eq:power_law_ratios} shows explicitly that $\tfz<\tcz$ in the presence of rotation. This implies that $\Tax$ transitions from $\Tsph$ to a higher value for $r\ll R_c$ due to compressive heating. Since the assumed $v_c$ is constant, this increase is logarithmic. Thus, the assumption of $\Tax=const.$ in the power law solution is correct up to small logarithmic variations.

The special value $F_0=3/2$ results in a zero flow time and infinite temperature. To understand this anomaly, we write the compressive heating rate to leading order
\begin{equation*}
    \dot{e}_{\rm CH}\propto \nabla\cdot\vec{v}\approx \frac{1}{r^2}\regd{}{r}\lbrac{r^2\vrax}-2\frac{\vrax}{r}F_\theta\approx \lbrac{\frac{3}{2}-F_0}\frac{\vrax}{r}.
\end{equation*}
Thus, the flow becomes incompressible for $F_0=3/2$, and compressive heating vanishes. In this case, no steady state power law solution can be found as there is no heating to balance the cooling.

Since $v_c^2/\csax^2\propto 1-2/3F_0$, we get
\begin{equation}\label{eq:power_law_temp}
\Tax = \Tsph\lbrac{1-2/3F_0}^{-1},
\end{equation}
and
\begin{equation*}
    \tfz =\tcsph\lbrac{1-\frac{2}{3}F_0}^{-1/2}\frac{\rhosph}{\rhoax},
\end{equation*}
where $\tcsph$ and $\rhosph$ are the cooling time and density in a spherical, non-rotating cooling flow with the same integrated $\dot{M}$ (eqn.~\ref{eq:spherical_solutions}), and we approximated $\Lambda\!\propto\! T^{-1/2}$, appropriate for $10^5\K\!<\!\Tax\!<\!10^7\K$. From mass conservation (eqn.~\ref{eq:mass_naa}), $\rhosph/\rhoax =\eta \vrax/\vrsph$, and we find that the radial velocity is
\begin{equation}
    \vrax = \lbrac{1-\frac{2}{3}F_0}^{1/4}\eta^{-1/2}\vrsph, 
\end{equation}
and the density is
\begin{equation}
    \rhoax = \rhorcf\equiv \lbrac{1-\frac{2}{3}F_0}^{-1/4}\eta^{-1/2}\rhosph,
\end{equation}
and since $\Pax\propto \rhoax\Tax$, we get
\begin{equation}
    \Pax = \lbrac{1-\frac{2}{3}F_0}^{-5/4}\eta^{-1/2}P_{\rm sph},
\end{equation}
where $P_{\rm sph} = 2/3\rhosph v_c^2$ is the pressure in the spherical solution with the same integrated $\dot{M}$.
For $F_0=1$, we get
\begin{equation}\label{eq:vr_cooling_flow}
    \vrax=\frac{r}{R_c}\vrsph\propto r^{1/2},
\end{equation}
\begin{equation}\label{eq:rho_cooling_flow}
    \rhoax = 1.8\frac{r}{R_c}\rhosph\propto r^{-1/2},
\end{equation}
and 
\begin{equation}\label{eq:P_cooling_flow}
    \Pax =  5.3\frac{r}{R_c}P_{\rm sph}\propto r^{-1/2}.
\end{equation}

\begin{figure}
    \centering
    \includegraphics[width=1.08\linewidth]{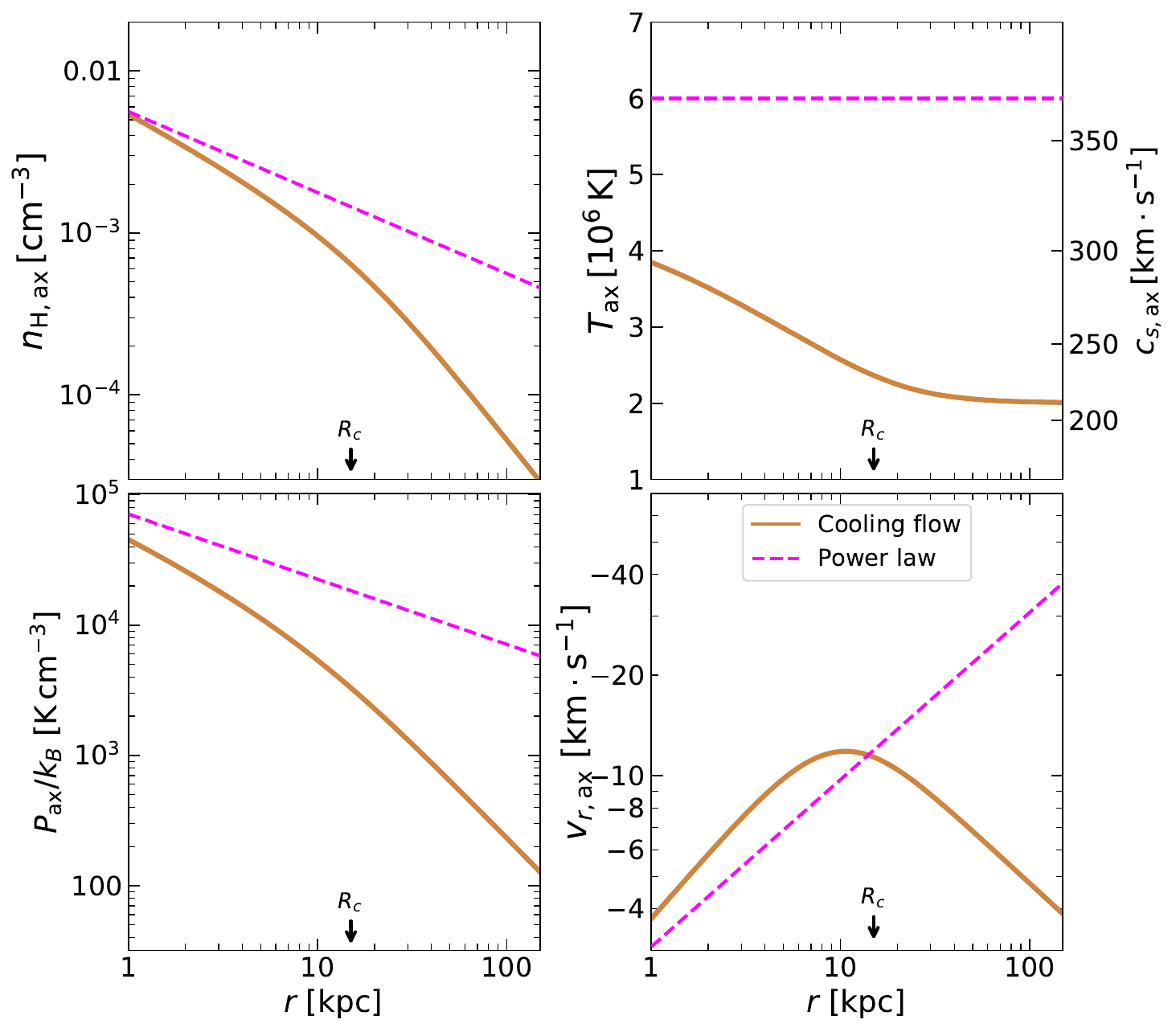}
    \caption{Similar to Fig. \ref{fig:analytic_vs_simulations}, except that the pressure profile is shown instead of the entropy. The gold solid curves correspond to the cooling flow solution of Section \ref{sec:analytic_results}. The magenta dashed lines show the power law solutions presented in this section. The full solution asymptotically approaches the analytic solution at $r\ll Rc$, up to a factor of $\sim\!2$ in $T$.}
    \label{fig:cold_disk_vs_zg}
\end{figure}

Fig. \ref{fig:cold_disk_vs_zg} shows a comparison between the cooling flow solution of Section \ref{sec:analytic_results} and the power law solutions. We can see the power law solution (magenta lines) approximate the cooling flow solution for the density, temperature and pressure well at $r\!\ll\!R_c$, with at most a factor of two deviation from the full solution. One can find next order corrections to the power law solutions at $r\ll R_c$ by assuming a varying temperature. The corrections are weak power laws, scaling roughly as $r^{0.26}$, and do not introduce qualitative changes to the power law solutions other than a slowly increasing temperature.
\subsection{Isentropic solutions}\label{sec:isentropic_solutions}
In the limit where $\tfz\ll\tcz$, the flow becomes isentropic, and eqn.~\eqref{eq:temperature_naa} can be integrated from $\rin$, using $\csax^2\propto \Tax$, which results in
\begin{equation}
    \Tax = T_{\rm in}\lbrac{1-(\gamma-1)\frac{\vc^2}{c_{\rm s,ax}^2}\ln\lbrac{\frac{r}{\rin}}}\equiv T_{\rm in}H(r),
\end{equation}
where $T_{\rm in}$ is the resulting temperature at $\rin$. Since the flow is isentropic, $\rhoax\propto \Tax^{1/(\gamma-1)}$, which yields
\begin{equation}\label{eq:isen_rho}
    \rhoax = \rhoin H(r)^{1/(\gamma-1)}.
\end{equation}
where $\rhoin$ is the resulting density at $\rin$. From mass conservation (eqn.~\ref{eq:mass_naa}), we get
\begin{equation}\label{eq:isen_vr}
    \vrax(r) = \frac{\dot{M}}{4\pi r^2\eta\rho} = v_{\rm in}\frac{\eta(\rin)}{\eta(r)}\lbrac{\frac{r}{\rin}}^{-2}H(r)^{-1/(\gamma-1)},
\end{equation}
where $\vin=\vrax(\rin)$ is the imposed inner velocity. Since $\rin\ll R_c$ and assuming $F_0=1$, we get from mass conservation that $\rhoin=(9/5)\dot{M}/4\pi R_c^2\vin$.

The flow remains isentropic up to a radius $\risen$. Before estimating it below, we can write the density and temperature by integrating from $\risen$, assuming $\Tax(\risen)\approx \Tax(\rout)$
\begin{equation}\label{eq:isen_T_out}
    \Tax = \Tax(\rout)\lbrac{1+(\gamma-1)\frac{\vc^2}{c_{\rm s,sph}^2}\ln\lbrac{\frac{\risen}{r}}}\equiv T_{\rm sph}G(r),
\end{equation}
\begin{equation}\label{eq:isen_rho_out}
    \rhoax = \rhoax(\rout)G(r)^{1/(\gamma-1)}.
\end{equation}

To estimate $\risen$, we require $\tfz(\risen)\approx \tcz(\risen)$. Assuming $\risen > R_c$, equating $\tfz$ with $\tcz$, using eqns.~\eqref{eq:isen_rho}-\eqref{eq:isen_vr}, results in the implicit equation for $\risen$
\begin{equation}\label{eq:full_risen}
    \risen^3H^{2/(\gamma-1)} = \frac{m_p}{\mu X^2(\gamma-1)}\frac{k_B\Tax(\rout)}{\Lambda}\frac{4\pi}{\dot{M}}\lbrac{\frac{5}{9}R_c^2}^2\vin^2,
\end{equation}
where we assumed $\Tax(\risen)\approx \Tax(\rout)$.
Since $H$ evolves only logarithmically, we can estimate $\risen$ by plugging $r=\rout$ into $H$, which is the maximum value for $\risen$. Using eqn.~\eqref{eq:isen_T_out} with $\rout=300\kpc$ and $\rin=1\kpc$ yields $H\approx 0.23$. We therefore get for $F_0=1$
\begin{equation}
\begin{split}
    \risen & \approx 95\kpc \lbrac{\frac{\vin}{40\kms}}^{2/3}\lbrac{\frac{H}{0.23}}^{-1}\\ 
    &\lbrac{\frac{\Tax(\rout)}{2\!\times\!10^6\K}}^{1/3}\lbrac{\frac{\dot{M}}{1\Msunyr}}^{-1/3}\lbrac{\frac{R_c}{15\kpc}}^{4/3}.
\end{split}
\end{equation}
In particular, $\risen\propto \vin^{2/3}$, thus larger velocities cause isentropic flow to larger radii. Solving for $\risen$ self consistently in eqn.~\eqref{eq:full_risen} yields a slightly smaller $\risen\approx 80\kpc$ for the same parameters.

\end{document}